\documentclass[a4paper,11pt]{article}
\usepackage{jheppub} 
\usepackage[T1]{fontenc} 
\usepackage{physics}
\RequirePackage{verbatim}
\NewDocumentCommand{\tens}{t_}
{%
	\IfBooleanTF{#1}
	{\tensop}
	{\otimes}%
}
\NewDocumentCommand{\tensop}{m}
{%
	\mathbin{\mathop{\otimes}\displaylimits_{#1}}%
}
\usepackage{graphicx}
\usepackage{wrapfig}
\usepackage{amsmath}
\usepackage{amsfonts}
\usepackage{amssymb}
\usepackage{subcaption}
\usepackage{dsfont}

\title{\textbf{\boldmath Unruh quantum Otto engine in the presence of a reflecting boundary}}

\author[a,1]{Arnab Mukherjee\note{Corresponding author.},}
\author[b]{Sunandan Gangopadhyay,}
\author[c]{A. S. Majumdar}

\affiliation{\textit{ S.N.~Bose National Centre for Basic Sciences, JD Block, Sector-III, Salt Lake, Kolkata 700106, India}}
\emailAdd{arnab.mukherjee@bose.res.in}
\emailAdd{sunandan.gangopadhyay@bose.res.in}
\emailAdd{archan@bose.res.in}

\abstract{\noindent We introduce a new model of relativistic quantum analogue of the classical Otto engine in the presence of a perfectly reflecting boundary.  A single qubit acts as the working substance interacting with a massless quantum scalar field, with the boundary obeying the Dirichlet condition.  The quantum vacuum serves as a thermal bath through the Unruh effect. We observe that the response function of the qubit gets significantly modified by the presence of 
the reflecting boundary. From the structure of the correlation function, we find that three different cases emerge, namely, the intermediate boundary regime, the near boundary regime, and the far boundary regime. As expected, the correlation in the far boundary regime approaches that of the Unruh quantum Otto engine (UQOE) when the reflecting boundary goes to infinity. The effect of the reflecting boundary is manifested through the  reduction of the critical excitation probability of the qubit and the work output of the engine.  
Inspite of the reduced work output, the efficiency of the engine remains 
unaltered even in the presence of the boundary.}

\begin{document} 
	\maketitle
	\flushbottom
	
\section{Introduction}\label{sec:Intro}
In recent years there has been an upsurge in interest in a field known as quantum thermodynamics which makes a connection between two fundamental physical theories, namely quantum mechanics and thermodynamics \cite{gemmer2009quantum, kosloff2013quantum, alicki2018introduction, deffner2019quantum}. A long standing question that exists is whether it is possible to derive the laws of thermodynamics from quantum principles. This has made quantum thermodynamics an active area of research \cite{vedral2002uniqueness, brandao2013resource}. Extensive studies on heat engines treated quantum mechanically \cite{rezek2006irreversible, wang2009performance, abe2011similarity, thomas2011coupled, kosloff2017quantum, agarwal2013quantum, rossnagel2013nano, azimi2014quantum, zhang2014quantum, ivanchenko2015quantum, de2019efficiency, camati2019coherence, del2022quantum}, have led to remarkable results and insights. By considering two level quantum systems, quantum versions of classical thermodynamic cycles have been proposed \cite{kieu2004second, kieu2006quantum, quan2007quantum, quan2009quantum, maruyama2009colloquium}. These are mainly quantum generalisations of the classical prototype of combustion engines. 

Studies on quantum analogues of classical heat engines have gained importance from the perspective of gravitational physics. Connections between relativistic quantum mechanics, thermodynamics and black hole physics has already been established in some seminal works \cite{bekenstein2020black, bardeen1973four, hawking1975particle, hawking1976black, unruh1976notes}. Therefore, it is quite natural to include relativistic notions in the domain of quantum thermodynamics. Relativistic extensions of  quantum thermodynamic engines have been carried out in \cite{papadatos2020relativistic, chattopadhyay2019relativistic, papadatos2021quantum, myers2021quantum}.

Investigations on field theoretic and relativistic phenomena in the presence of static or accelerating reflecting boundaries is an emerging topic in recent times.  Accelerating mirrors can be considered as an analogue of the dynamical Casimir effect in (1+1) dimensions \cite{2010grae.book.....H, moore1970quantum}, a prototype for black hole evaporation \cite{fulling1976radiation, davies1977radiation, juarez2018quantum, cong2020effects, wilson2020tidal}. It 
has been observed that there is a strong connection between accelerating mirrors and black hole physics \cite{good2017horizonless, good2019information, fernandez2022duality, myrzakul2021cghs}.

Innovation in nanofabrication techniques \cite{PhysRevLett.104.203603, PhysRevLett.109.033603} have enhanced  the scope of experimental realization of atomic excitations in nanoscale waveguides \cite{corzo2019waveguide}, through trapped atoms in optical nanofibers \cite{doi:10.1126/science.1237125, SOLANO2017439}. These avenues open up the possibilities of exploration of fundamental quantum optical aspects like atom-photon lattices \cite{RevModPhys.90.031002}. Studies on relativistic quantum phenomena in superconducting circuits \cite{PhysRevLett.110.113602, PhysRevB.92.064501}, and secure quantum communication over long-distances \cite{huang2019protection, huang2020deterministic, PhysRevD.104.105020, hengl2009directed}, show the importance of reflecting  boundaries which also play significant physical role in the context of atom-field interaction \cite {PhysRevD.104.124001, refId0}, holographic entanglement entropy \cite{akal2021holographic}, and quantum entanglement \cite{PhysRevD.75.104014, zhou2013boundary, PhysRevD.98.025001, liu2021entanglement}.

Motivated by the importance of the reflecting boundary in the context of superconducting circuits, quantum communication, atom-field interaction and quantum entanglement, in this paper we aim to investigate the effect of a
reflecting boundary in case of quantum heat engines. To begin this investigation, at first we introduce a new model for the relativistic quantum analogue of the classical Otto engine \cite{arias2018unruh, gray2018scalar, xu2020unruh, kane2021entangled, barman2022constructing}. Here we consider a single qubit (Unruh-DeWitt detector) acting as the working substance, interacting linearly with a massless quantum scalar field in the presence of a perfectly reflecting boundary which obeys the Dirichlet boundary condition. The quantum vacuum serves as a thermal bath through the Unruh effect \cite{unruh1976notes}. 

The importance of carrying out this investigation in the presence of a reflecting boundary lies in its relevance to cavity quantum electrodynamics which is a thrust area of fundamental research and has practical applications \cite{haroche2006exploring}. In particular, techniques of quantum electrodynamics in a cavity can be applied to investigate the Unruh-Davies effect inside cavities \cite{PhysRevLett.91.243004, PhysRevA.74.023807}. Cavity quantum electrodynamical setups are also important in superconducting circuits which can implement large acceleration \cite{PhysRevB.92.064501}. The presence of boundaries have also been incorporated in theoretical investigations of radiative processes of entangled atoms \cite{arias2016boundary}.

One can understand the effect of the reflecting boundary from two perspectives.  From a 
physical point of view, Poincare symmetry is broken by the presence of the boundary and thus, spectral density of the field is also modified, which accordingly alters the dynamics of the detector-field system \cite{zhou2013boundary}. On the other hand, from another technical viewpoint, exploiting the symmetry in the model, one can solve the problem of the qubit interacting with a boundary  modified field by using the method of image charge problem. It is well known that the correlation function between the scalar fields commonly known as the Wightman function gets significantly modified by the presence of the reflecting boundary \cite{rizzuto2007casimir}. 

Since the response function of the qubit only depends on this correlation function of the scalar field \cite{sachs2017entanglement, louko2016unruh, takagi1986vacuum, PhysRevD.93.024019}, so boundary effects must be captured in the response function of the qubit. The presence of reflecting boundary affects the quantum fields \cite{astrakhantsev2018massive}, the correlation function between the fields and the rate of
spontaneous emission of excited atoms \cite{Purcell1995}. Therefore, one would expect modifications in the response function and transition probability of the qubit. 

In the present work we explore how such modifications in the response
function affect the transition probability and the amount of work extraction from the quantum vacuum fluctuation of the massless scalar field. We choose
the system parameters in the range of current experiemntal values in domain
of superconducting circuits where large atomic acceleration can be realized \cite{garcia2017entanglement}. Based on such a parameter  range, it is
possible to demarcate  three different regimes, namely, the near boundary regime,
the intermediate boundary regime, and the far boundary regime, as we show in
our analysis. We study the
behaviour of the above observable quantities in each case as a function of the qubit
acceleration, as well as the initial qubit state parameters, and the
bath temperatures. We observe
several interesting variations in the observable properties corresponding to
the three different regimes. We find that though the reflecting boundary
adversely impacts the work output, the   efficiency of the heat engine 
remains unaltered. 

The paper is organised as follows. In section \ref{sec:QOE}, we recapitulate certain basic definitions of quantum thermodynamics and quantum Otto cycles. In section \ref{sec:UQOEM}, a new form of relativistic extension of the quantum Otto cycle is introduced. Sections \ref{sec:qu-vac int} and \ref{sec:WhF} are devoted to study how a qubit interacts with the quantum vacuum leading to the modification of the correlation function  in the presence of the reflecting boundary. In section \ref{sec:RF}, the response function, which is a key factor for calculating the transition probability, is evaluated. In section \ref{sec:Th-Analysis} the thermodynamical analysis of the heat engine is performed. Section \ref{sec:Effects on transition probability and work output} contains a detailed analysis of the 
effects due to the reflecting boundary on the transition probability, work output and efficiency. Finally, we conclude in section \ref{sec:Con}.


\section{Quantum Otto cycle}\label{sec:QOE}
Thermodynamic cycles provide foundational working principles of the heat engines \cite{cengel2011thermodynamics}. To begin our discussion, we start by reviewing the basic features of the quantum otto cycle (QOC).  In the quantum counterpart of the classical Otto cycle for a qubit \cite{kieu2004second, kieu2006quantum} the energy gap between the two levels are not fixed and can be adjusted by an external (weak electric or magnetic) stimulation without hampering the state of the system. Considering a qubit having a ground state $\vert g\rangle$, and an excited state $\vert e\rangle$, described by a density matrix $\rho(t)$ and associated time dependent Hamiltonian $\mathcal{H}(t)$, the average energy of the system $\langle E(t)\rangle = \text{Tr}[\rho(t)\mathcal{H}(t)]$ satisfies the following equation \cite{balian2006microphysics}
\begin{eqnarray}
\partial_{t}\langle E(t)\rangle=\text{Tr}[\partial_{t}\rho(t)\mathcal{H}(t)]+\text{Tr}[\rho(t)\partial_{t}\mathcal{H}(t)]\,\,.
\end{eqnarray}
This equation is identical with the first law thermodynamics and can be treated as its quantum version where the first term and second term on the right hand side can be described as the change in the internal state and population of the system and the external shifting of the energy levels of the system, respectively. Therfore, the following identifications can be made 
\begin{eqnarray}
\langle Q\rangle&=&\int_0^{\mathcal{T}}\, dt\,\text{Tr}\left[\frac{\partial \rho(t)}{\partial t}\mathcal{H}(t)\right]\label{avrgQ}\\
\langle W\rangle&=&\int_0^{\mathcal{T}}\, dt\,\text{Tr}\left[\rho(t)\frac{\partial \mathcal{H}(t)}{\partial t}\right]\label{avrgW}
\end{eqnarray}
where $\langle Q\rangle$ is the average heat transfer to the system and $\langle W\rangle$ is the work done on the system over the interaction time 
$\mathcal{T}$.

\begin{figure}[!htbp]
\centering
\includegraphics[scale=0.55]{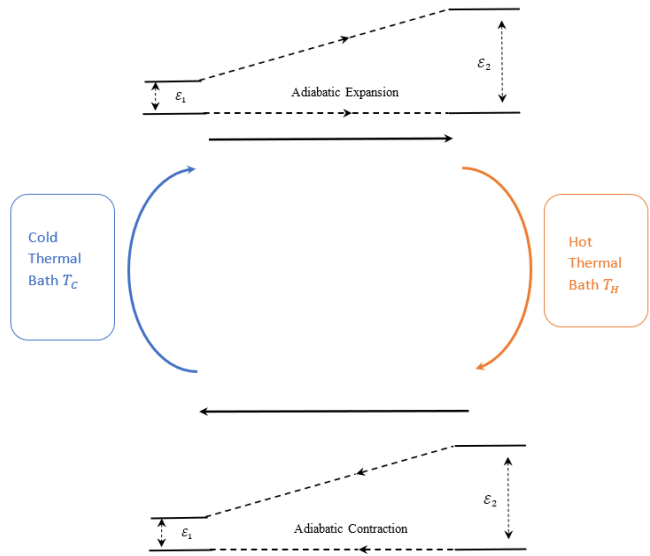}
\caption{Thermodynamic cycle of Quantum Otto Engine.}
\label{fig:QOE}
\end{figure}

Considering the ground state and excited state energy of the qubit being zero and $\mathcal{E}_1$ respectively, the Hamiltonian of the qubit turns out to be $H=\mathcal{E}_1 \vert e\rangle\langle e\vert$. We also consider that the Otto cycle begins with the system in an initial state $\rho_0=p\vert e\rangle\langle e\vert+(1-p)\vert g\rangle\langle g\vert$. Thermodynamical steps of the QOC are as follows
\begin{enumerate}
\item In the first step, without changing form of the initial state the system undergoes an adiabatic expansion of the energy gap $\mathcal{E}_1$ to $\mathcal{E}_2$. In this step, there is no heat exchange with the environment but some work is done.
\item In the second step, the qubit $\rho_0$ is attached with a thermal bath at temperature $T_H$. After the interaction for the time $\mathcal{T}_2$, heat is exchanged and the initial state of the system is changed to, $\rho=(p+\delta p_H)\vert e\rangle\langle e\vert+(1-p-\delta p_H)\vert g\rangle\langle g\vert$ where $\delta p_H$ is the transition probability between two levels of the qubit due to the first interaction with the thermal bath. In this step, the total amount of work done is zero.
\item In the third step, keeping the state fixed at $\rho$ the system undergoes an adiabatic compression and the energy level of the qubit is reduced from $\mathcal{E}_2$ to $\mathcal{E}_1$. In this step also, no heat is exchanged with the environment but a sufficient amount of work is done. This is the power stroke of the cycle where the system performs work.
\item In the final step, the system is attached with a thermal bath having a lesser temperature compared to that of the previous bath, $T_C < T_H$. After interaction time $\mathcal{T}_1$, the final state of the qubit becomes $\rho_f=(p+\delta p_H+\delta p_C)\vert e\rangle\langle e\vert+(1-p-\delta p_H-\delta p_C)\vert g\rangle\langle g\vert$, where $\delta p_C$ is the transition probability between two levels of the qubit due to the second interaction with the thermal bath. In this step also, no work is done. 
\end{enumerate}
For completing the cycle, we must have $\delta p_H+\delta p_C=0$. It is worth noting here that using the von Neumann entropy $S = -k_{B}\text{Tr}[\rho \log\rho]$ and the equilibrium Boltzmann distribution $\rho = \text{exp}(- \beta \mathcal{H})/Z$ where $Z$ is the partition function, one can show that $T \delta S = \text{Tr}[\delta \rho \mathcal{H}] = \delta Q$, which shows the compatibility of the quantum thermodynamics domain with the usual classical thermodynamics.


\section{Unruh quantum Otto engine in the presence of a reflecting boundary}\label{sec:UQOEM}
\begin{figure}[!h]
\begin{minipage}{.5\textwidth}
\centering
\includegraphics[width=1.1\linewidth]{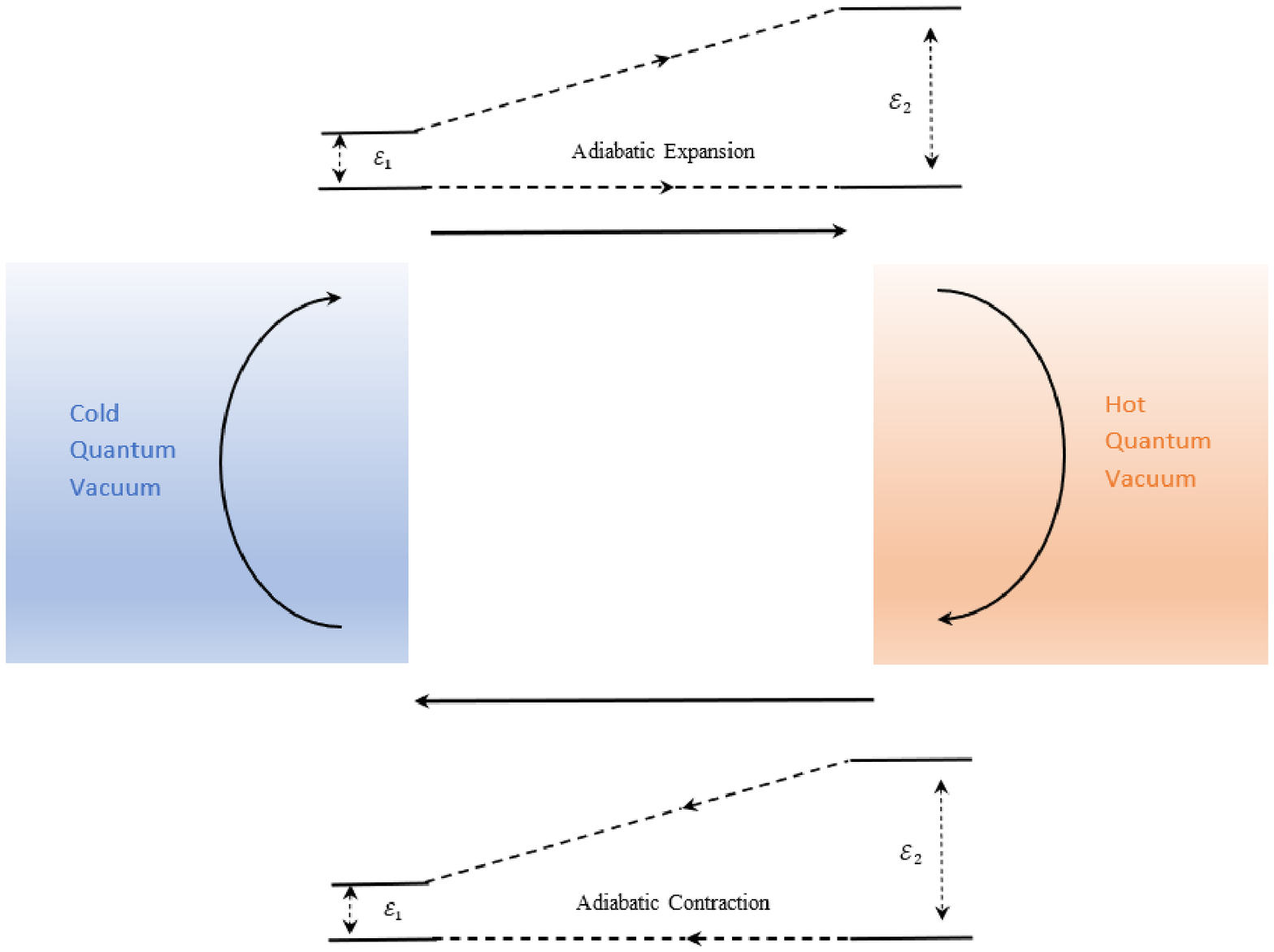}
\subcaption{Thermodynamic cycle}
\label{fig:UQOE}
\end{minipage}
\hspace{1cm}
\begin{minipage}{.5\textwidth}
\centering
\includegraphics[width=1.1\linewidth]{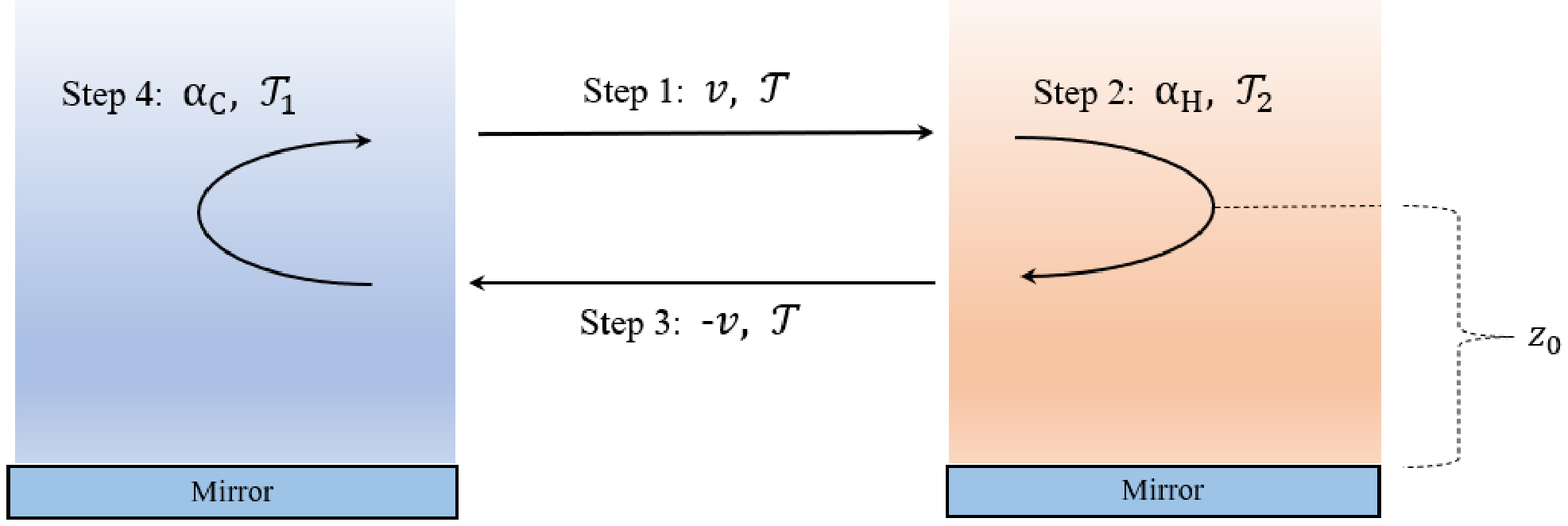}
\subcaption{Kinematic cycle}
\label{fig:MUQOE}
\end{minipage}
\caption{Different cycles of UQOE with a single reflecting boundary.}
\end{figure}

 Extension of the quantum Otto engine (QOE) to the relativistic domain has been
  made ealier by exploiting the notion of Unruh effect \cite{arias2018unruh}. Here our aim is to see that how a Unruh quantum Otto engine (UQOE) behaves in the presence of a reflecting boundary. To address this point, we introduce a single perfectly reflecting boundary. Consider that the qubit is at a distance of $z_0$ from the boundary and accelerates along the $x$ direction as depicted in Figure \ref{fig:MUQOE}.
  
From the thermodynamical cycle of UQOE in the presence of a single reflecting boundary (Figure \ref{fig:UQOE}), steps 2 and 4 of the QOC cold and hot temperatures will correspond to different accelerations $\alpha_H$ and $\alpha_C$. As the temperature of the vacuum is proportional of the particle's acceleration,  we must have $\alpha_H > \alpha_C$ for one vacuum to have higher temperature than the other, i.e., $T_H > T_C$, and thus transfer work from the vacuum to the system. We will assume in the other two steps (step 1 and step 3) that the qubit  travels at constant velocity and can be isolated from the quantum field vacuum.
Apart from this we also require that the kinematic cycle of the qubit (Figure \ref{fig:MUQOE}) is closed.
Steps of the kinematic cycle of UQOE with a single reflecting boundary are as follows.
\begin{enumerate}
\item The qubit  moves at constant velocity $v$ for time $\mathcal{T}$, during which the energy gap expands from $\mathcal{E}_1$ to $\mathcal{E}_2$. This step is similar to the adiabetic expansion.
\item In this step, the qubit undergoes a constant acceleration $\alpha_H$ over the interaction time $\mathcal{T}_2$. During this interaction velocity of the qubit changes from $v$ to $-v$ and the quantum vacuum acts as a hot thermal reservoir. 
\item The qubit moves at constant velocity $-v$ for time $\mathcal{T}$, during which the energy gap reduces from $\mathcal{E}_2$ to $\mathcal{E}_1$. This step is similar to the adiabetic contraction.
\item In this step, the qubit undergoes a constant acceleration $\alpha_C$ over the interaction time $\mathcal{T}_1$. During this interaction the quantum vacuum serves as a cold thermal reservoir. The velocity of the qubit changes from $-v$ to $v$ and it returns to its initial state.
\end{enumerate}
Here $\mathcal{T}_1$ and $\mathcal{T}_2$ are the time in which the qubit accelerates and interacts with the quantum vacuum (the velocity of the qubit changes from $v$ to $-v$, and $-v$ to $v$, respectively). For closing the kinematical cycle and returing the qubit to it's original state, $\mathcal{T}_1$ and $\mathcal{T}_2$ must have some fixed value.

In the relativistic picture it is well known that a constantly accelerated observer moving with a proper acceleration $\alpha$ in proper time $\tau$ has a hyperbolic worldline $x^{\mu}(\tau)=(t(\tau),\textbf{x}(\tau))$ where
\begin{eqnarray}
t(\tau)=\frac{c}{\alpha}\sinh(\frac{\alpha\tau}{c})\qquad
x(\tau)=\frac{c^2}{\alpha}\cosh(\frac{\alpha \tau}{c})\,.
\end{eqnarray}
onsidering above worldline for a qubit, the qubit velocity $v$ turns out to be
\begin{eqnarray}
v=c\tanh(\frac{\alpha \tau}{c})
\end{eqnarray}
Therefore, the time taken by the qubit to reach the velocity $v$ from $\tau=0$ is
\begin{eqnarray}
\tau = \frac{c}{\alpha} \tanh^{-1}(\frac{v}{c})
=\frac{c}{\alpha} \tanh^{-1}(\beta)
\end{eqnarray}
Now, if we consider that at time $t=-\tau$ and $t=\tau$ the velocity of the qubit is $-v$ and $v$ respectively, then it takes the time $2\tau=\frac{2c}{\alpha} \tanh^{-1}(\beta)$ to change the velocity from $-v$ to $v$ with a constant acceleration $\alpha$.
Hence, in second and fourth step in the thermodynamic as well as the kinematic cycle, the qubit must accelerate for the following times
\begin{eqnarray}
\mathcal{T}_2= \frac{2c}{\alpha_H} \tanh^{-1}(\beta)\\
\mathcal{T}_1= \frac{2c}{\alpha_C} \tanh^{-1}(\beta)
\end{eqnarray}
with accelerations $\alpha_H$ and $\alpha_C$, respectively. From here we can conclude that the vacuum  behaves as a hot or cold reservoir over these amounts of finite interaction times.


\section{Qubit-vacuum interaction}\label{sec:qu-vac int}

We consider a qubit interacting with a real massless scalar field $\phi[x(\tau)]$ linearly, in the presence of a reflecting boundary where the distance between the qubit and the boundary is $z_0$. The Hamiltonian that describes the qubit-field interacting system in the instantaneous inertial frame of the qubit reads \cite{rizzuto2007casimir}
\begin{eqnarray}
\mathcal{H}=\mathcal{H}_{qubit}+\mathcal{H}_{field}+\mathcal{H}_{int}
\end{eqnarray}
where
\begin{eqnarray}
\mathcal{H}_{qubit}&=&\mathcal{E}\vert e\rangle\langle e\vert\\
\mathcal{H}_{field}&=&\int \text{d}^{3}k\,\mathcal{E}(k)\,{b}^{\dagger}(\mathbf{k})b(\mathbf{k})\\
\mathcal{H}_{int}&=&\lambda\,m(\tau)\,\phi[x(\tau)].
\end{eqnarray}
Here $b(\mathbf{k})$ and $b^{\dagger}(\mathbf{k})$ are the bosonic operators of the scalar field \cite{peskin2018introduction}, $\lambda$ is a weak qubit field coupling constant having the dimension $[M^{3/2}L^{5/2}T^{-1}]$ and $m(\tau)$ is the monopole operator of the qubit.
It may be noted that in this model we consider that the qubit acts as a point particle and the total Hamiltonian of the system is written in the interaction picture so that we can use the free mode expansion of the massless scalar field $\phi[x(\tau)]$. The monopole operator $m(\tau)$ can be written in this picture as
\begin{eqnarray}
m(\tau)=e^{+\frac{i\mathcal{E}\tau}{\hbar}}\vert e\rangle\langle g\vert+e^{-\frac{i\mathcal{E}\tau}{\hbar}}\vert g\rangle\langle e\vert= \begin{pmatrix}
0&e^{+i\mathcal{E}\tau/\hbar}\\e^{-i\mathcal{E}\tau/\hbar}&0
\end{pmatrix}.
\end{eqnarray}
\subsection{Evolution of the qubit}
Let us consider that the initial density matrix of the qubit is given by
\begin{eqnarray}
\rho_0=p\vert e\rangle\langle e\vert+(1-p)\vert g\rangle\langle g\vert= \begin{pmatrix}
p&0\\0&1-p
\end{pmatrix}
\end{eqnarray}
where $p$ is the probability for the qubit remaining in the excited state.
In an inertial frame, the initial density matrix of the scalar field having the vacuum state $\vert 0\rangle$ is given by
\begin{eqnarray}
\rho_{field}=\vert 0\rangle\langle 0\vert.
\end{eqnarray}
Therefore, the initial state of the interaction between qubit and the quantum vacuum can be taken as
\begin{eqnarray}
\varrho_0=\rho_0\otimes\rho_{field}\,\,.
\end{eqnarray}
Now, using the time evolution operator $U_\mathcal{T}$, we can write down the density matrix of the interaction over a period $\mathcal{T}$ as
\begin{eqnarray}
\varrho_\mathcal{T}=U_\mathcal{T}\,\varrho_0\, U^{\dagger}_\mathcal{T}\,\,.\label{evolve-rho}
\end{eqnarray}
The time evolution operator $U_\mathcal{T}$ satisfies the evolution equation
\begin{eqnarray}
i\hbar\frac{\partial U_\mathcal{T}}{\partial \tau}=\mathcal{H}_{int}(\tau)\,U_\mathcal{T}\label{q-Li0}
\end{eqnarray}
whose solution is given by the Dyson series \cite{peskin2018introduction}
\begin{eqnarray}
U_\mathcal{T}=  \mathds{1} -\frac{i}{\hbar}\int_{t_0}^{t}\text{d}\tau\,\mathcal{H}_{int}(\tau)-\frac{1}{2\hbar^2} \int_{t_0}^{t}\text{d}\tau\int_{t_0}^{t}\text{d}\tau'\,\hat{T}\{\mathcal{H}_{int}(\tau)\,\mathcal{H}_{int}(\tau')\}+\mathcal{O}(\lambda^3)\label{dyson}
\end{eqnarray}
where $\hat{T}$ is the time ordering operator defined as 
\begin{eqnarray}
\hat{T}\{A(t)B(t')\}=\Theta(t-t')A(t)B(t')+\Theta(t'-t)B(t')A(t).
\end{eqnarray}
Substituting eq.(\ref{dyson}) into eq.(\ref{evolve-rho}), we get
\begin{eqnarray}
\varrho_\mathcal{T}&=&\varrho_0 - \frac{i}{\hbar}\int_{t_0}^{t}\text{d}\tau\,[\mathcal{H}_{int}(\tau),\varrho_0]-\frac{1}{2\hbar^2} \int_{t_0}^{t}\text{d}\tau\int_{t_0}^{t}\text{d}\tau'\,\hat{T}\{[\mathcal{H}_{int}(\tau),\,[\mathcal{H}_{int}(\tau'),\varrho_0]]\}+\mathcal{O}(\lambda^3)\,.\nonumber\\
&\,& \label{rho-mixed}
\end{eqnarray}
Our primary concern is the evolution of the qubit in the interaction picture. As we are not interested about the evolution of the vacuum state, hence after taking a partial trace over the field degrees of freedom to extract the evolution of the qubit state, we get
\begin{eqnarray}
\rho_\mathcal{T}=\text{Tr}_\text{field}[\varrho_\mathcal{T}]\,\,.
\end{eqnarray}
Calculating the partial trace in eq.(\ref{rho-mixed}) term by term, we see that the zeroth order term is simply the initial state of the qubit, the first order term vanishes as the vacuum expectation value of the field operator vanishes, while the second order term gives the main contribution to the evolving state. Hence, after the interaction time $\mathcal{T}$, the density matrix of the qubit becomes
\begin{eqnarray}
\rho_\mathcal{T}=\begin{pmatrix}
p+\delta p_\mathcal{T}&0\\0&1-p-\delta p_\mathcal{T}
\end{pmatrix}+\mathcal{O}(\lambda^4)
\end{eqnarray}
where $\delta p_\mathcal{T}$ is the transition probability between the levels of the qubit occurring due to quantum vacuum fluctuations. Therefore, transition probability can be recast as
\begin{eqnarray}
\delta p_\mathcal{T}=\frac{\lambda^2}{\hbar^2} \int_{-\tilde{t}}^{t}\text{d}\tau\int_{-\tilde{t}}^{t}\text{d}\tau'\,((1-p)\,e^{-i\mathcal{E}(\Delta \tau)/\hbar}-p\,e^{+i\mathcal{E}(\Delta \tau)/\hbar})\mathcal{G}^{+}(\tau,\tau')\label{deltap}
\end{eqnarray}
where $\mathcal{G}^{+}(\tau,\tau')$ is known as the positive frequency Wightman function \cite{birrell1984quantum} given by
\begin{eqnarray}
\mathcal{G}^{+}(\tau,\tau')=\langle 0\vert\, \phi[x(\tau)]\phi[x(\tau')]\,\vert 0\rangle\,.
\end{eqnarray}
Introducing a switching function $\varpi_{\mathcal{T}}(\tau)$ and taking the Markovian approximation, eq.(\ref{deltap}) becomes \cite{arias2018unruh, gray2018scalar}
\begin{eqnarray}
\delta p_\mathcal{T}&=&\frac{\lambda^2}{\hbar^2} \int_{-\infty}^{+\infty}\text{d}\tau\int_{-\infty}^{+\infty}\text{d}\tau'\,\varpi_{\mathcal{T}}(\tau)\,\varpi_{\mathcal{T}}(\tau')\,((1-p)\,e^{-i\mathcal{E}(\Delta \tau)/\hbar}-p\,e^{+i\mathcal{E}(\Delta \tau)/\hbar})\mathcal{G}^{+}(\tau,\tau')\nonumber\\&=&\lambda^2[(1-p)\,\mathcal{F}(\mathcal{E},\mathcal{T})-p\,\mathcal{F}(-\mathcal{E},\mathcal{T})].\label{deltap1}
\end{eqnarray}
The response function of the qubit can be defined as 
\begin{eqnarray}
\mathcal{F}(\mathcal{E},\mathcal{T})=\frac{1}{\hbar^2}\int_{-\infty}^{+\infty}\text{d}\tau \int_{-\infty}^{+\infty}\text{d}\tau '\varpi_{\mathcal{T}}(\tau)\,\varpi_{\mathcal{T}}(\tau')\,\mathcal{G}^{+}(\tau,\tau')\,\text{e}^{-i\,\mathcal{E}(\Delta \tau)/\hbar}.
\end{eqnarray}


\section{Vacuum correlation function}\label{sec:WhF}
 In the presence of a reflecting boundary, the Wightman function is the sum of the empty space part and a part that depends on the presence of the mirror \cite{rizzuto2007casimir}. We can write it as \cite{birrell1984quantum}
\begin{eqnarray}
\mathcal{G}^{+}(\tau ,\tau ')=-\frac{\hbar}{4\pi ^2 c}\left[ \frac{1}{(c\,\Delta t-i\eta)^2-(x-x')^2-(y-y')^2-(z-z')^2}\right.\nonumber\\
\left. \,\,\,\,-\frac{1}{(c\,\Delta t-i\eta)^2-(x-x')^2-(y-y')^2-(z+z')^2}\right]\label{gt}
\end{eqnarray}
where $\Delta t$ is the difference between qubit coordinates $t(\tau)$ at two different proper times and $\eta$ is a  small parameter.
In the laboratory frame the trajectory of an uniformly accelerating qubit along the $x$ direction at a distance $z_0$ from the reflecting boundary reads,
\begin{eqnarray}
t(\tau)&=&\frac{c}{\alpha}\sinh\left(\frac{\alpha \tau}{c}\right), \hspace{1em}
x(\tau)=\frac{c^2}{\alpha}\cosh\left(\frac{\alpha \tau}{c}\right)\nonumber\\
y(\tau)&=& 0, \hspace{1em}
z(\tau)= z_0 \,\,. \label{tra1}
\end{eqnarray}
Here $\alpha$ is the proper acceleration and $\tau$ is the proper time of the qubit.
Now, using  eqs.(\ref{tra1}), we get
\begin{eqnarray}
c\,\Delta t &=&c\,( t(\tau)-t(\tau'))\nonumber\\
&=& \frac{c^2}{\alpha}\left[\sinh\left(\frac{\alpha \tau}{c}\right)-\sinh\left(\frac{\alpha \tau'}{c}\right)\right]\,\,.
\end{eqnarray}
Similarly, we also have the following relation
\begin{eqnarray}
\Delta x=\frac{c^2}{\alpha}\left[\cosh\left(\frac{\alpha \tau}{c}\right)-\cosh\left(\frac{\alpha \tau'}{c}\right)\right],\,\,\Delta y=0\,,\,\,\Delta z=0\,\,.
\end{eqnarray}
Using the above two results and keeping terms upto $\mathcal{O}(\eta)$, we get
\begin{eqnarray}
(c\,\Delta t-i\eta)^2-(\Delta x)^2&=&\left[\frac{c^2}{\alpha}\left\{\sinh\left(\frac{\alpha \tau}{c}\right)-\sinh\left(\frac{\alpha \tau'}{c}\right)\right\}-i\eta\right]^2-\left[\frac{c^2}{\alpha}\left\{\cosh\left(\frac{\alpha \tau}{c}\right)-\cosh\left(\frac{\alpha \tau'}{c}\right)\right\}\right]^2\nonumber\\
&=&-\frac{2c^4}{\alpha^2}+\frac{2c^4}{\alpha^2}\left\{\cosh\left(\frac{\alpha \tau}{c}\right)\,\cosh\left(\frac{\alpha \tau'}{c}\right)-\sinh\left(\frac{\alpha \tau}{c}\right)\,\sinh\left(\frac{\alpha \tau'}{c}\right)\right\}-i\eta\nonumber\\
&=&-\frac{2c^4}{\alpha^2}+\frac{2c^4}{\alpha^2}\left\{\cosh\left(\frac{\alpha (\Delta\tau)}{c}\right)-i\eta\right\}\nonumber\\
&=&-\frac{2c^4}{\alpha^2}\left[1-\cosh\left(\frac{\alpha (\Delta\tau)}{c}-i\eta\right)\right]\nonumber\\
&=&\frac{4c^4}{\alpha^2}\sinh^2\left(\frac{\alpha (\Delta\tau)}{2c}-i\eta\right)\label{gp}
\end{eqnarray}
where $\Delta\tau$ is the difference between two different proper times $\tau$ and $\tau'$.
 In a similar way, it can be shown that
\begin{eqnarray}
(c\,\Delta t-i\eta)^2-(x-x')^2-(y-y')^2-(z+z')^2&=&\frac{4c^4}{\alpha ^2}\left[\sinh ^2\left(\frac{\alpha (\Delta\tau)}{2c}-i\eta\right)-\frac{z^2_0\alpha ^2 }{c^4}\right]\,\,.\nonumber\\
&\,& \label{gm}
\end{eqnarray}
Therefore, after using eq.(s)(\ref{gp}, \ref{gm}) in eq.(\ref{gt}), the Wightman function in the presence of a reflecting boundary takes the form
\begin{eqnarray}
\mathcal{G}^{+}(\tau ,\tau ')=-\frac{\hbar\,\alpha ^2}{16\pi ^2c^5}\left[ \frac{1}{\sinh ^2\left(\frac{\alpha (\Delta\tau)}{2c}-i\eta\right)}-\frac{1}{\sinh ^2\left(\frac{\alpha (\Delta\tau)}{2c}-i\eta\right)-\frac{z^2_0\alpha ^2 }{c^4}}\right]\,\,.\label{gt1}
\end{eqnarray}


\section{Evaluation of the response function}\label{sec:RF}
In this section, we proceed to calculate the response function.
From the time evolution of the qubit eq.\eqref{deltap1}, the transition probability between the energy levels of the qubit can be written as
\begin{eqnarray}
\delta p=\lambda^2[(1-p)\,\mathcal{F}(\mathcal{E},\mathcal{T})-p\,\mathcal{F}(-\mathcal{E},\mathcal{T})]\label{deltapp}
\end{eqnarray}
where
\begin{eqnarray}
\mathcal{F}(\mathcal{E},\mathcal{T})=\frac{1}{\hbar^2}\int_{-\infty}^{+\infty}\text{d}\tau \int_{-\infty}^{+\infty}\text{d}\tau '\varpi_{\mathcal{T}}(\tau)\,\varpi_{\mathcal{T}}(\tau')\,\mathcal{G}^{+}(\tau,\tau')\,\text{e}^{-i\,\mathcal{E}(\Delta\tau)/\hbar}\,\,.
\end{eqnarray}
Following \cite{arias2018unruh, gray2018scalar}, we  now consider a Lorentzian switching function 
\begin{eqnarray}
\varpi_{\mathcal{T}}(\tau)=\frac{(\mathcal{T}/2)^2}{\tau^2+(\mathcal{T}/2)^2}\,\,.\label{lsf0}
\end{eqnarray}
Changing the variables from $(\tau,\tau')\rightarrow(m,n)$ by the transformations $m=\tau-\tau'$ and $n=\tau+\tau'$, we can recast the above equation as
\begin{eqnarray}
\mathcal{F}(\mathcal{E},\mathcal{T})&=&\frac{1}{2\hbar^2}\int_{-\infty}^{+\infty}\text{dm}\left(\mathcal{G}^{+}(\tau,\tau')\,\text{e}^{-i\,\mathcal{E}m/\hbar} \int_{-\infty}^{+\infty}\text{dn}\,\varpi_{\mathcal{T}}((m+n)/2)\,\varpi_{\mathcal{T}}((m-n)/2)\right)\,\,.\nonumber\\ 
&\,& \label{rf}
\end{eqnarray}
The advantage of the Lorentzian regulator is that it enables us to extend the integration to the complex plane and use the residue theorem since $\varpi_{\mathcal{T}}(z)\rightarrow0$ for $\vert z\vert\rightarrow\infty$ for all $z\in\mathbb{C}$.
Putting eq.(\ref{lsf0}) in the integral containing swtiching functions, we get
\begin{eqnarray}
\int_{-\infty}^{+\infty}\text{dn}\,\varpi_{\mathcal{T}}((m+n)/2)\,\varpi_{\mathcal{T}}((m-n)/2)&=&\int_{-\infty}^{+\infty}\text{dn}\frac{\mathcal{T}^4}{[(n+m)^2+\mathcal{T}^2][(n-m)^2+\mathcal{T}^2]}\,\,.\nonumber\\
&\,& \label{lsf}
\end{eqnarray}

\noindent \begin{figure}[!h]
\centering
\includegraphics[scale=0.6]{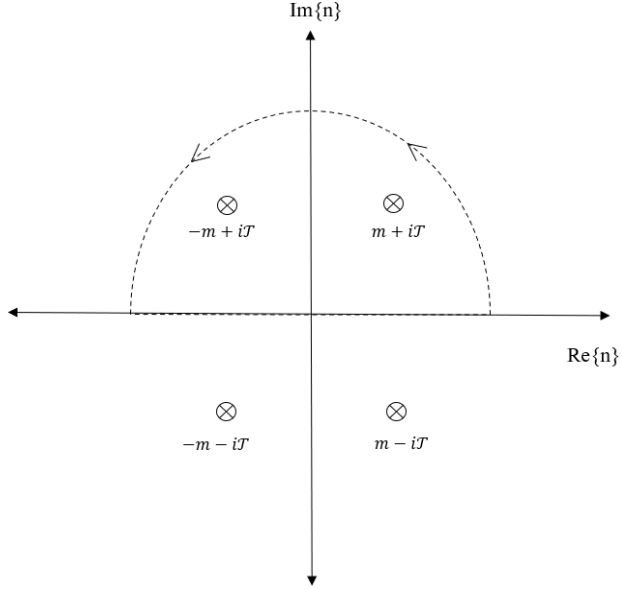}
\caption{Pole structures of the integral (\ref{lsf}).}
\label{fig:pole_n}
\end{figure}

\noindent The above integral can be evaluated by employing the method of contour integral \cite{freitag2009complex}. Figure \ref{fig:pole_n} shows the pole structure of eq.(\ref{lsf}). Considering the contour in the upper half plane we see that the pole $n=m+i\mathcal{T}$ and $n=-m+i\mathcal{T}$ lies inside the contour. Therefore, residues at the point $n=m+i\mathcal{T}$ and $n=-m+i\mathcal{T}$ read
\begin{eqnarray}
R_1=\frac{\mathcal{T}^4}{8im\mathcal{T}(m+i\mathcal{T})}\\
R_2=\frac{\mathcal{T}^4}{8im\mathcal{T}(m-i\mathcal{T})}\,\,.
\end{eqnarray}
Hence, we get
\begin{eqnarray}
\int_{-\infty}^{+\infty}\text{dn}\frac{\mathcal{T}^4}{[(n+m)^2+\mathcal{T}^2][(n-m)^2+\mathcal{T}^2]}&=&2\pi i\,[R_1+R_2]\nonumber\\
&=&2\pi i\left[\frac{\mathcal{T}^4}{8im\mathcal{T}(m+i\mathcal{T})}+\frac{\mathcal{T}^4}{8im\mathcal{T}(m-i\mathcal{T})}\right]\nonumber\\
&=&\frac{\pi \mathcal{T}^3}{2}\frac{1}{(m^2+\mathcal{T}^2)}\,\,.\label{lsf1}
\end{eqnarray}
Using the result of the Lorentzian switching function integral eq.(\ref{lsf1}) in eq.(\ref{rf}), the response function simplifies to
\begin{eqnarray}
\mathcal{F}(\mathcal{E},\mathcal{T})=\frac{\pi \mathcal{T}^3}{4\hbar^2}\int_{-\infty}^{+\infty}\text{dm}\,\frac{\mathcal{G}^{+}(\tau,\tau')}{m^2+\mathcal{T}^2}\,\text{e}^{-i\,\mathcal{E}m/\hbar}\,\,.\label{rf1}
\end{eqnarray}
Now, incorporating the Wightman function in the above equation, we can evaluate the response function for the following scenarios.

\subsection{Evaluation of transition probability}
Using eq.(\ref{gt1}) in eq.(\ref{rf1}), we have
\begin{eqnarray}
\mathcal{F}(\mathcal{E},\mathcal{T})=\mathcal{F}_1(\mathcal{E},\mathcal{T})+\mathcal{F}_2(\mathcal{E},\mathcal{T})\label{F0}
\end{eqnarray}
where
\begin{eqnarray}
\mathcal{F}_1(\mathcal{E},\mathcal{T})&=&-\frac{\pi \mathcal{T}^3}{4\hbar c}\frac{\alpha ^2}{16\pi ^2 c^4}\int_{-\infty}^{+\infty}\text{dm}\,\frac{\text{e}^{-i\,\mathcal{E}m/\hbar}}{m^2+\mathcal{T}^2}\,\frac{1}{\sinh ^2\left(\frac{\alpha m}{2c}-i\eta\right)}\\
\mathcal{F}_2(\mathcal{E},\mathcal{T})&=&\frac{\pi \mathcal{T}^3}{4\hbar c}\frac{\alpha ^2}{16\pi ^2 c^4}\int_{-\infty}^{+\infty}\text{dm}\,\frac{\text{e}^{-i\,\mathcal{E}m/\hbar}}{m^2+\mathcal{T}^2}\,\frac{1}{\sinh ^2\left(\frac{\alpha m}{2c}-i\eta\right)-\frac{z^2_0\alpha ^2 }{c^4}}\,\,.
\end{eqnarray}
Defining dimensionless variables 
\begin{eqnarray}
z&=&z_0\alpha/c^2\,, \hspace{1cm}\mu=\mathcal{E}c/(\hbar \alpha)\nonumber\\ 
\nu&=&\alpha \mathcal{T}/c\,, \hspace{1cm}\xi=\alpha m/c\label{dlvari}
\end{eqnarray}
and recasting the above integrals in terms of these dimensionless variables, we get
\begin{eqnarray}
\mathcal{F}_{1}(\mu, \nu) = -\frac{\nu^3}{64\pi\hbar c^3}\int_{-\infty}^{+\infty}\text{d}\xi\,\frac{\text{e}^{-i\mu \xi}}{\xi ^2+\nu ^2}\,\frac{1}{\sinh ^2 \left(\frac{\xi}{2}-i\eta\right)}\label{F1}
\end{eqnarray}
and
\begin{eqnarray}
\mathcal{F}_{2}(\mu, \nu) = \frac{\nu^3}{64\pi\hbar c^3}\int_{-\infty}^{+\infty}\text{d}\xi\,\frac{\text{e}^{-i\mu \xi}}{\xi ^2+\nu ^2}\,\frac{1}{\big[\sinh ^2 \left(\frac{\xi}{2}-i\eta\right)-z^2\big]}\,\,.\label{F2}
\end{eqnarray}
To carry out the integrals in eq(s). (\ref{F1}, \ref{F2}), we use the method of contour integration.

\subsubsection{Calculation of $\mathcal{F}_{1}(\mu, \nu)$}
Using the series representation \cite{Gradshteyn:1702455}
\begin{eqnarray}
\csch^2\left[\frac{\xi}{2} -i\eta\right]=\displaystyle\sum_{k=-\infty}^{\infty}\frac{4}{(\xi -2i\eta -2i\pi k)^2}=\displaystyle\sum_{k=-\infty}^{\infty}\frac{4}{(\xi -i\epsilon -2i\pi k)^2}\label{csch2-series}
\end{eqnarray}
where $\epsilon=2\eta$, in eq.(\ref{F1}), we get
\begin{eqnarray}
\mathcal{F}_{1}(\mu, \nu)=\frac{1}{\hbar c^3} \left[\mathcal{I}_0+\displaystyle\sum_{k=1}^{\infty}\mathcal{I}_k\right]\label{F11}
\end{eqnarray}
where $I_0$ and $I_k$ are given by the integrals
\begin{eqnarray}
\mathcal{I}_{0}&=&-\frac{\nu ^3}{16\pi }\int_{-\infty}^{+\infty}\text{d}\xi\,\frac{\text{e}^{-i\mu \xi}}{\xi ^2+\nu ^2}\frac{1}{(\xi-i\epsilon)^2}\label{FI0}\\
\mathcal{I}_{k}&=&-\frac{\nu ^3}{16\pi }\int_{-\infty}^{+\infty}\text{d}\xi\,\frac{\text{e}^{-i\mu \xi}}{\xi ^2+\nu ^2}\left\{\frac{1}{(\xi-2i\pi k)^2}+\frac{1}{(\xi+2i\pi k)^2}\right\}\,\,.\label{FIk}
\end{eqnarray}
\begin{figure}[!h]
\centering
\includegraphics[scale=0.6]{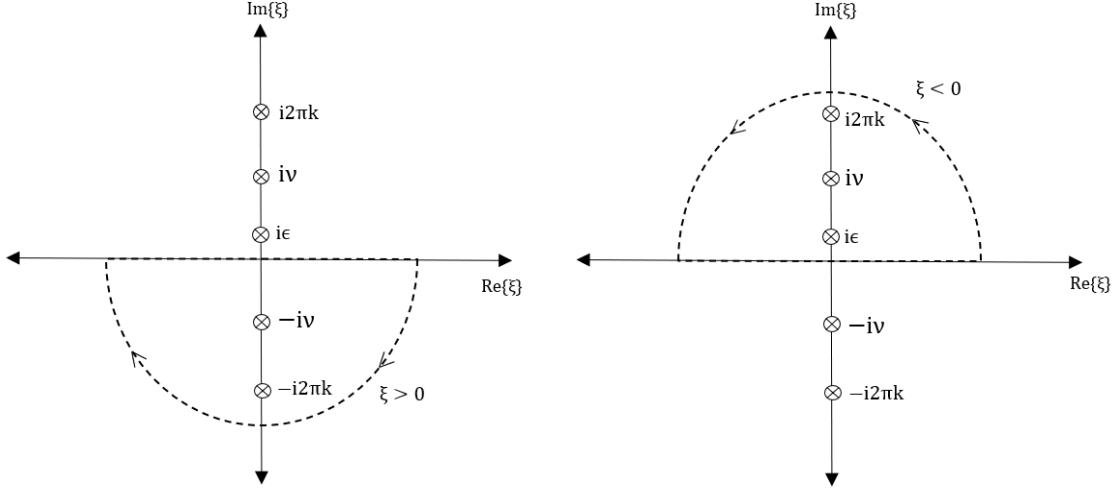}
\caption{Pole structures of the integrals (\ref{FI0}) and (\ref{FIk}).}
\label{fig:poles}
\end{figure}
For carrying out the integral $\mathcal{I}_0$, we consider the range $\mu<0$ and close the contour in the upper half plane and find one first order pole at $\xi=i\nu$ and one second order pole at $\xi=i\epsilon$.\\
Calculating the residues for all poles and taking the limit $\epsilon\rightarrow0$, we get  
\begin{eqnarray}
R_1=-e^{\mu\nu}/(2i\nu^3),\hspace{1cm}R_2=-i\mu/\nu^2\,\,.
\end{eqnarray}
Hence, we have
\begin{eqnarray}
\mathcal{I}_0&=&-\frac{\nu^3}{16\pi}\Big[2\pi i(R_1+R_2)\Big]\nonumber\\
&=&-\frac{\nu^3}{16\pi}\left[-\frac{\pi}{\nu^3}e^{\mu\nu}+2\pi\mu\frac{1}{\nu^2}\right]\,\,.
\end{eqnarray}
Similarly, when $\mu>0$, the contour has to be closed in the lower half plane and $\xi$ will only pick a first order pole at, $\xi=-i\nu$. Calculating the residue for the pole and taking the limit $\epsilon\rightarrow0$, we get  
\begin{eqnarray}
R_3=e^{-\mu\nu}/(2i\nu^3).
\end{eqnarray}
Hence, 
\begin{eqnarray}
\mathcal{I}_0&=&-\frac{\nu^3}{16\pi}\Big[-2\pi i(R_3)\Big]\nonumber\\
&=&-\frac{\nu^3}{16\pi}\left[\frac{\pi}{\nu^3}e^{\mu\nu}\right]\,.\label{FI01}
\end{eqnarray}
Therefore, considering both the region $\mu<0$ and $\mu>0$, we can write
\begin{eqnarray}
\mathcal{I}_0=\frac{1}{16}\left[e^{-\vert\mu\vert\nu}+2\vert\mu\vert\nu\Theta(-\mu)\right]\,
\end{eqnarray}
with
\begin{eqnarray}
\Theta(-\mu)&=&1, \,\,\,\text{when}\,\,\mu<0\nonumber\\
&=&0,\,\,\,\text{otherwise}\,.\nonumber
\end{eqnarray}
We now proceed to evaluate the integral $\mathcal{I}_{k}$ given in eq.\eqref{FIk}. Just like the previous case, here also at first we consider the range $\mu<0$ and close the contour in the upper half plane and find one first order pole at $\xi=i\nu$ and one second order pole at $\xi=2\pi ik$. Now, calculating the residues for all poles, taking the limit $\epsilon\rightarrow0$ and considering both the range $\mu<0$ and $\mu>0$, the final result of the integral in eq.(\ref{FIk}) reads
\begin{eqnarray}
\mathcal{I}_k&=&\frac{\nu^2\,e^{-\vert\mu\vert\nu}}{16}\left[\frac{1}{(\nu-2\pi k)^2}+\frac{1}{(\nu+2\pi k)^2}\right]+\frac{\nu^2}{64\pi^2}e^{-2\pi k\vert\mu\vert}\left[2\pi\vert\mu\vert\left\{\frac{1}{(k+\frac{\nu}{2\pi})}-\frac{1}{(k-\frac{\nu}{2\pi})}\right\}\right.\nonumber\\
&+&\left.\frac{1}{(k+\frac{\nu}{2\pi})^2}-\frac{1}{(k-\frac{\nu}{2\pi})^2}\right].\label{FIk1}
\end{eqnarray}
We now employ the Lerch-Hurwitz transcendental function \cite{ferreira2004asymptotic},
\begin{eqnarray}
\Phi (z,n,a)=\displaystyle\sum_{k=0}^{\infty}\frac{z^k}{(k+a)^n}\,\,.\label{Lerch}
\end{eqnarray}
Using the above definition of Lerch-Hurwitz transcendental function and the series representation of $\sin^2(\nu/2)$,
and taking the summation over $k$ in the eq. (\ref{FIk1}), we finally get
\begin{eqnarray}
\displaystyle\sum_{k=1}^{\infty}\mathcal{I}_k=\frac{e^{-\vert\mu\vert\nu}}{16}\left[\frac{(\nu/2)^2}{\sin^2(\nu/2)}-1\right]+\frac{\nu^2\,e^{-2\pi\vert\mu\vert}}{64\pi^2}\left[2\pi\vert\mu\vert\Delta\Phi(\mu,1,\nu)+\Delta\Phi(\mu,2,\nu)\right]
\end{eqnarray}
where $\Delta\Phi(\mu,n,\nu)$ is defined as
\begin{eqnarray}
\Delta\Phi(\mu,n,\nu)=\Phi\left(e^{-2\pi\vert\mu\vert},n,1+\frac{\nu}{2\pi}\right)-\Phi\left(e^{-2\pi\vert\mu\vert},n,1-\frac{\nu}{2\pi}\right)\,\,.
\end{eqnarray}
Collecting all these results and substituting them in eq. (\ref{F11}), we obtain
\begin{eqnarray}
\mathcal{F}_1(\mu,\nu)&=&\frac{1}{\hbar c^3}\bigg[\frac{1}{16}\left(2\vert\mu\vert\nu\Theta(-\mu)+\frac{(\nu/2)^2}{\sin^2(\nu/2)}e^{-\vert\mu\vert\nu}\right)+\frac{\nu^2\,e^{-2\pi\vert\mu\vert}}{64\pi^2}\Big[2\pi\vert\mu\vert\Delta\Phi(\mu,1,\nu)+\Delta\Phi(\mu,2,\nu)\Big]\bigg]\,\,.\nonumber\\
&\,&\label{finalF1}
\end{eqnarray}

\subsubsection{Calculation of $\mathcal{F}_{2}(\mu, \nu)$}
From eq. (\ref{F2}), one can find that the poles are situated at
\begin{eqnarray}
\xi=\pm i\nu,\hspace{0.5cm}\text{and}\hspace{0.5cm}\xi= \xi^{\pm}\equiv i\epsilon\pm 2\sinh^{-1}(z).
\end{eqnarray}
All the poles are first order in nature. Now considering the range $\mu<0$ and closing the contour in the upper half plane, we find that the two first order poles at $\xi=i\nu$ and $\xi=\xi^{+}$ lie inside the contour. Calculating the residue at $\xi=i\nu$ and taking the limit $\epsilon\rightarrow0$, we get
\begin{eqnarray}
R_1=-\frac{e^{\mu\nu}}{2i\nu}\left[\frac{1}{\sin^2(\nu/2)+z^2}\right].
\end{eqnarray}
Now from eq.(\ref{F2}), we can define the function under the integral sign $f(\xi)$ as
\begin{eqnarray}
f(\xi)=\frac{g(\xi)}{h(\xi)}
\end{eqnarray}
where,
\begin{eqnarray}
g(\xi)&=&\exp(-i\mu\xi)\\
h(\xi)&=&(\xi ^2+\nu ^2)\bigg[\sinh ^2 \left(\frac{\xi-i\epsilon}{2}\right)-z^2\bigg]\,\,.
\end{eqnarray}
Calculating the values of $h(\xi)$ and the first derivative of $h(\xi)$ at the point $\xi=\xi^{+}$, we get
\begin{eqnarray}
h(\xi^{+})=0\,,\hspace{0.5in} h^{\prime}(\xi^{+})\neq 0.
\end{eqnarray}
Since at the point $\xi=\xi^{+}$, $h(\xi^{+})=0$ but $h^{\prime}(\xi^{+})\neq 0$, therefore residue at the point $\xi=\xi^{+}$ can be written as
\begin{eqnarray}
R_2=\frac{g(\xi^{+})}{h^{\prime}(\xi^{+})}\,\,.
\end{eqnarray}
Calculating this and taking the limit $\epsilon\rightarrow0$, we find
\begin{eqnarray}
R_2=\frac{\exp(-2i\mu\sinh^{-1}(z))}{(4\sinh^{-2}(z)+\nu^2)}\frac{1}{z\sqrt{1+z^2}}\,\,.
\end{eqnarray}
This then gives
\begin{eqnarray}
\mathcal{F}_2(\mu,\nu)&=&\frac{\nu^3}{64\pi\hbar c^3}\times 2\pi i (R_1+R_2)\nonumber\\
&=&\frac{\nu^3}{64\hbar c^3}\Bigg[-\frac{e^{\mu\nu}}{\nu}\left[\frac{1}{\sin^2(\nu/2)+z^2}\right]+\frac{2i\exp(-2i\mu\sinh^{-1}(z))}{(4\sinh^{-2}(z)+\nu^2)}\frac{1}{z\sqrt{1+z^2}} \Bigg]\,\,.\nonumber\\
&\,&
\end{eqnarray}
Evaluating eq.(\ref{F2}) by considering the range $\mu>0$, we see that the relevant poles are at $\xi=-i\nu$ and $\xi=\xi^{-}$. Hence considering both the range $\mu<0$ and $\mu>0$, we find 
\begin{eqnarray}
\mathcal{F}_2(\mu,\nu)&=&\frac{\nu^3}{64\hbar c^3}\Bigg[-\frac{e^{-\vert\mu\vert\nu}}{\nu}\left[\frac{1}{\sin^2(\nu/2)+z^2}\right]+\frac{2i\exp(2i\vert\mu\vert\sinh^{-1}(z))}{(4\sinh^{-2}(z)+\nu^2)}\frac{1}{z\sqrt{1+z^2}} \Bigg]\,\,.\nonumber\\
&\,&
\end{eqnarray}
The real part of $\mathcal{F}_2(\mu,\nu)$ therefore becomes
\begin{eqnarray}
\text{Real}\,\mathcal{F}_2(\mu,\nu)=-\frac{1}{16\hbar c^3}\frac{(\nu/2)^2 e^{-\vert\mu\vert\nu}}{[\sin^2(\nu/2)+z^2]}\,\,.\label{finalF2}
\end{eqnarray}\\
Substituting eq.(s)(\ref{finalF1}, \ref{finalF2}) in eq.(\ref{F0}), the complete response function turns out to be
\begin{eqnarray}
\mathcal{F}(\mu,\nu)&=&\frac{1}{\hbar c^3}\Bigg[\frac{1}{16}\bigg[2\vert\mu\vert\nu\Theta(-\mu)+(\nu/2)^2e^{-\vert\mu\vert\nu}\left\{\frac{1}{\sin^2(\nu/2)}-\frac{1}{[\sin^2(\nu/2)+z^2]}\right\}\bigg]\Bigg.\nonumber\\
&+&\Bigg.\frac{\nu^2\,e^{-2\pi\vert\mu\vert}}{64\pi^2}\bigg[2\pi\vert\mu\vert\Delta\Phi(\mu,1,\nu)+\Delta\Phi(\mu,2,\nu)\bigg]\Bigg]\,\,.\label{finalF0}
\end{eqnarray}

Introducing a new parameter known as the reduced acceleration $a=1/\mu$, where $\mu$ is defined in eq.\eqref{dlvari}, we can recast the response function (in terms of the reduced acceleration and the ratio of the qubit's velocity to that of light in vacuum) as 
\begin{eqnarray}
\mathcal{F}\left(\frac{1}{a},\, 2\tanh^{-1}(\beta)\right)&=&\frac{1}{\hbar c^3}\left[\frac{1}{4\vert a\vert}\tanh^{-1}(\beta)\Theta\left(-\frac{1}{a}\right)+\frac{e^{-2\,\frac{1}{\vert a\vert}\tanh^{-1}(\beta)}\tanh^{-2}(\beta)}{16}\left[\frac{1}{\sin^2\{\tanh^{-1}(\beta)\}}\right.\right.\nonumber\\
&-&\left.\frac{1}{[\sin^2\{\tanh^{-1}(\beta)\}+z^2]}\right]+\frac{\tanh^{-2}(\beta)}{16\pi^2}e^{-2\pi\frac{1}{\vert a\vert}}\left[2\pi\frac{1}{\vert a\vert}\Delta\Phi\left(\frac{1}{a},\,1,\,2\tanh^{-1}(\beta)\right)\right.\nonumber\\
&+&\left.\left.\Delta\Phi\left(\frac{1}{a},\,2,\,2\tanh^{-1}(\beta)\right)\right]\right].\label{com-rf00}
\end{eqnarray}
Defining $\Delta\mathcal{F}\left(\frac{1}{a},\, 2\tanh^{-1}(\beta)\right)=-\mathcal{F}\left(\frac{1}{a},\, 2\tanh^{-1}(\beta)\right)+\mathcal{F}\left(-\frac{1}{a},\, 2\tanh^{-1}(\beta)\right)$, we can rewrite eq.(\ref{deltapp}) as
\begin{eqnarray}
\delta p(a,\,p,\,\beta)&=&\lambda_0^2\left[(1-2p)\,\hbar c^3 \mathcal{F}\left(\frac{1}{a},\, 2\tanh^{-1}(\beta)\right)-p\,\hbar c^3\Delta\mathcal{F}\left(\frac{1}{a},\, 2\tanh^{-1}(\beta)\right)\right]\hspace{1cm}\label{deltap0}
\end{eqnarray}
where from dimensional analysis we fix a dimensionless parameter $\lambda_0=\sqrt{\lambda^2 c/\hbar^3}$, and $\Delta\mathcal{F}\left(\frac{1}{a},\, 2\tanh^{-1}(\beta)\right)$ is given by
where $\Delta\mathcal{F}\left(\frac{1}{a},\, 2\tanh^{-1}(\beta)\right)$ is given by
\begin{eqnarray}
\Delta\mathcal{F}\left(\frac{1}{a},\, 2\tanh^{-1}(\beta)\right)=\frac{1}{\hbar c^3}\left[\frac{1}{4\vert a\vert}\tanh^{-1}(\beta)\right].\label{deltap00}
\end{eqnarray}

The structure of the response function enables us to demarcate two limiting
cases through the condition $\sin^{2}(\nu/2)\sim z^{2}$ which defines the intermediate boundary regime. Two other regimes emerge from this definition as we shall see in the subsequent subsections.

\subsection{Evaluation of transition probability in the near boundary regime}

In the near boundary regime, we have $z^2<<\sin^2(\nu/2)$. Hence, carrying out a series expansion of eq.\eqref{finalF0} for small $z$, we obtain
\begin{eqnarray}
\mathcal{F}(\mu,\nu)&=&\frac{1}{\hbar c^3}\Bigg[\frac{1}{16}\bigg[2\vert\mu\vert\nu\Theta(-\mu)+(\nu/2)^2e^{-\vert\mu\vert\nu}\left\{\frac{z^2}{\sin^4(\nu/2)}\right\}\bigg]\Bigg.\nonumber\\
&+&\Bigg.\frac{\nu^2\,e^{-2\pi\vert\mu\vert}}{64\pi^2}\bigg[2\pi\vert\mu\vert\Delta\Phi(\mu,1,\nu)+\Delta\Phi(\mu,2,\nu)\bigg]\Bigg]\,\,.\label{finalFF1}
\end{eqnarray}
Recasting the response function in terms of the reduced acceleration (a) and the ratio of the qubit's velocity to that of light in vacuum, we get
\begin{eqnarray}
\mathcal{F}\left(\frac{1}{a},\, 2\tanh^{-1}(\beta)\right)&=&\frac{1}{\hbar c^3}\Bigg[\frac{1}{4\vert a\vert}\tanh^{-1}(\beta)\Theta\left(-\frac{1}{a}\right)+\frac{e^{-2\,\frac{1}{\vert a\vert}\tanh^{-1}(\beta)}\tanh^{-2}(\beta)}{16}\left[\frac{z^2}{\sin^4\{\tanh^{-1}(\beta)\}}\right]\Bigg.\nonumber\\
&+&\Bigg.\frac{\tanh^{-2}(\beta)}{16\pi^2}e^{-2\pi\frac{1}{\vert a\vert}}\left[2\pi\frac{1}{\vert a\vert}\Delta\Phi\left(\frac{1}{a},\,1,\,2\tanh^{-1}(\beta)\right)+\Delta\Phi\left(\frac{1}{a},\,2,\,2\tanh^{-1}(\beta)\right)\right]\Bigg]\nonumber\\
&\,&\label{com-rf001}
\end{eqnarray}
where $\Delta\mathcal{F}\left(\frac{1}{a},\, 2\tanh^{-1}(\beta)\right)$ is given by
\begin{eqnarray}
\Delta\mathcal{F}\left(\frac{1}{a},\, 2\tanh^{-1}(\beta)\right)=\frac{1}{\hbar c^3}\left[\frac{1}{4\vert a\vert}\tanh^{-1}(\beta)\right].\label{deltap001}
\end{eqnarray}

\subsection{Evaluation of transition probability in the far boundary regime}
In the far boundary regime, we have $z^2>>\sin^2(\nu/2)$. Hence, carrying out a series expansion of eq.\eqref{finalF0} for large $z$, we obtain
\begin{eqnarray}
\mathcal{F}(\mu,\nu)&=&\frac{1}{\hbar c^3}\left[\frac{1}{8}\vert\mu\vert\nu\,\Theta(-\mu)+\frac{e^{-\vert\mu\vert\nu}}{16}\left[\frac{(\nu/2)^2}{\sin^2(\nu/2)}-\frac{\nu^2}{4z^2}+\frac{\nu^2}{4z^4}\sin^2(\nu/2)\right]\right.\nonumber\\
&+&\left.\frac{\nu^2}{64\pi^2}e^{-2\pi\vert\mu\vert}\left[2\pi\vert\mu\vert\Delta\Phi(\mu,1,\nu)+\Delta\Phi(\mu,2,\nu)\right]\right]\,.\label{com-rf21}
\end{eqnarray}
Now recasting the response function in terms of the reduced acceleration $a$, and the ratio of the qubit's velocity to that of light in vacuum, we get
\begin{eqnarray}
\mathcal{F}\left(\frac{1}{a},\, 2\tanh^{-1}(\beta)\right)&=&\frac{1}{\hbar c^3}\left[\frac{1}{4\vert a\vert}\tanh^{-1}(\beta)\Theta\left(-\frac{1}{a}\right)+\frac{e^{-2\,\frac{1}{\vert a\vert}\tanh^{-1}(\beta)}}{16}\left[\frac{\tanh^{-2}(\beta)}{\sin^2\{\tanh^{-1}(\beta)\}}\right.\right.\nonumber\\
&-&\left.\frac{\tanh^{-2}(\beta)}{z^2}\left(1-\frac{\sin^2\{\tanh^{-1}(\beta)\}}{z^2}\right)\right]+\frac{\tanh^{-2}(\beta)}{16\pi^2}e^{-2\pi\frac{1}{\vert a\vert}}\nonumber\\
&\times &\left.\left[2\pi\frac{1}{\vert a\vert}\Delta\Phi\left(\frac{1}{a},\,1,\,2\tanh^{-1}(\beta)\right)+\Delta\Phi\left(\frac{1}{a},\,2,\,2\tanh^{-1}(\beta)\right)\right]\right]\nonumber\\
&\,&\label{com-rf22}
\end{eqnarray}
where $\Delta\mathcal{F}\left(\frac{1}{a},\, 2\tanh^{-1}(\beta)\right)$ is given by
\begin{eqnarray}
\Delta\mathcal{F}\left(\frac{1}{a},\, 2\tanh^{-1}(\beta)\right)=\frac{1}{\hbar c^3}\left[\frac{1}{4\vert a\vert}\tanh^{-1}(\beta)\right].\label{deltap22}
\end{eqnarray}


\section{Analysis of thermodynamical steps}\label{sec:Th-Analysis}
In this section, we will analyse each thermodynamical step of the UQOE in the presence of a reflecting boundary and calculate the amount of heat exchanged between the qubit and the quantum vacuum and the amount of work done by the qubit.
\subsection{Adiabatic expansion}
In this step, the form of the initial state of the qubit $\rho_0=p\vert e\rangle\langle e\vert+(1-p)\vert g\rangle\langle g\vert$ remains fixed and the energy gap between the energy levels changes from $\mathcal{E}_1$ to a higher value $\mathcal{E}_2$ over a time $\mathcal{T}$. The time-dependent Hamiltonian of the  qubit is given by
\begin{eqnarray}
\mathcal{H}(t)=\mathcal{E}(t)\vert e\rangle \langle e \vert.
\end{eqnarray}
Using the definition of the average heat transfer eq.\eqref{avrgQ}, we find that this step is purely adiabatic, i.e., 
\begin{eqnarray}
\langle Q_1\rangle=\int_0^{\mathcal{T}}\, dt\,\text{Tr}\left[\frac{\partial \rho_0}{\partial t}\mathcal{H}(t)\right]=0.
\end{eqnarray}
In a similar way, using the expression of average work done eq.\eqref{avrgW}, we  find that there is a positive work done on the system, given by
\begin{eqnarray}
\langle W_1\rangle &=&\int_0^{\mathcal{T}}\, dt\,\text{Tr}\left[\rho_0 \frac{\partial \mathcal{H}(t)}{\partial t}\right]\nonumber\\
&=&\int_0^{\mathcal{T}}\, d\mathcal{E}\,\text{Tr}\left[\rho_0\vert e\rangle \langle e\vert\right]\nonumber\\
&=&p(\mathcal{E}_2 - \mathcal{E}_1).
\end{eqnarray}

\subsection{Contact with the hot vacuum}
In this step, the Hamiltonian of the system is fixed at a constant value $\mathcal{H}=\mathcal{E}_2 \vert e\rangle\langle e\vert$. The qubit accelerates from $v$ to $-v$ over the interval $\mathcal{T}_2$ and interacts with the background quantum field. During this time the qubit's state evolves through the interaction with the background quantum field as shown in section \ref{sec:qu-vac int} and takes the form 
\begin{eqnarray}
\rho_{\mathcal{T}_2}=\rho_0+\delta p_H \sigma_3
\end{eqnarray}
where $\sigma_3=\vert e\rangle\langle e\vert-\vert g\rangle\langle g\vert$ and $\delta p_H=\delta p_{\mathcal{T}_2}$.
No work is done in this step due to the constant value of the qubit Hamiltonian and hence, we get
\begin{eqnarray}
\langle W_2\rangle=\int_0^{\mathcal{T}_2}\, dt\,\text{Tr}\left[\rho(t)\frac{\partial \mathcal{H}(t)}{\partial t}\right]=0.
\end{eqnarray}
On the other hand, the system absorbs heat from the vacuum, given by
\begin{eqnarray}
\langle Q_2\rangle &=& \int_0^{\mathcal{T}_2}\, dt\,\text{Tr}\left[\frac{\partial \rho(t)}{\partial t}\mathcal{H}\right]
=\int_0^{\mathcal{T}_2}\, dt\,\text{Tr}\left[\frac{\partial \delta p(t)}{\partial t}\mathcal{E}_2 \sigma_3\vert e\rangle\langle e\vert\right]\nonumber\\
&=&\mathcal{E}_2\int_0^{\mathcal{T}_2}\text{Tr}\left[\partial \delta p(t) \sigma_3\vert e\rangle\langle e\vert\right]=\mathcal{E}_2\text{Tr}\left[\delta p_H \sigma_3\vert e\rangle\langle e\vert\right]\nonumber\\
&=&\mathcal{E}_2 \delta p_H.
\end{eqnarray}

\subsection{Adiabatic contraction}
In this step, the qubit travels at velocity $-v$ and the state $\rho$ is held fixed at $\rho=\rho_0+\delta p_H \sigma_3$ as the energy gap is reduced from $\mathcal{E}_2$ to $\mathcal{E}_1$. Just like the adiabatic expansion, no heat is exchanged, and we have
\begin{eqnarray}
\langle Q_3\rangle=0
\end{eqnarray}
and the value of work done is
\begin{eqnarray}
\langle W_3\rangle=-(\mathcal{E}_2-\mathcal{E}_1)(p+\delta p_H)\,.
\end{eqnarray}

\subsection{Contact with the cold vacuum}
In the final step, the Hamiltonian of the system is again fixed at another constant value $\mathcal{H}=\mathcal{E}_1 \vert e\rangle\langle e\vert$. The qubit accelerates from $-v$ to $+v$ over the interval $\mathcal{T}_1$ and interacts with the background quantum field. Therefore, just like the hot vacuum case the state of the qubit evolves and takes the form
\begin{eqnarray}
\rho_{\mathcal{T}_1}=\rho_1+\delta p_C \sigma_3
\end{eqnarray}
where $\rho_1=p'\vert e\rangle\langle e\vert+(1-p')\vert g\rangle\langle g\vert$, $\delta p_C=\delta p_{\mathcal{T}_1}$ and $p'=p+\delta p_H$. Here also we get no work done
\begin{eqnarray}
\langle W_4\rangle=0
\end{eqnarray}
and the average heat transfer is
\begin{eqnarray}
\langle Q_4\rangle=\mathcal{E}_1 \delta p_C.
\end{eqnarray}

\subsection{Completing the cycle}
From the above analysis we have already calculated and got the amount of heat exchanged and work done in each step of the thermodynamical cycle. Now, for returning the qubit to its initial state and completing the cycle, we have to impose the  condition  $\delta p_H+\delta p_C=0$.
The total amount of heat transfer and the net work done by the cycle is then
\begin{eqnarray}
\langle W_{\text{tot}}\rangle &=&\langle W_{1}\rangle +\langle W_{3}\rangle =-(\mathcal{E}_2-\mathcal{E}_1)\,\delta p_H\label{netw}\\
\langle Q_{\text{tot}}\rangle &=&\langle Q_{2}\rangle +\langle Q_{4}\rangle =(\mathcal{E}_2-\mathcal{E}_1)\,\delta p_H\label{neth}
\end{eqnarray}
which obeys the conservation of energy as
\begin{eqnarray}
\langle W_{\text{tot}}\rangle +\langle Q_{\text{tot}}\rangle =0\,.
\end{eqnarray} 
In case of the quantum thermal engine (QOE) discussed in section \ref{sec:QOE}, after exchanging the heat with the hot and cold reservoirs, the qubit state satisfies
\begin{eqnarray}
p+\delta p_H=\text{Tr}[\vert e\rangle\langle e\vert\rho]=1/(1+exp(\mathcal{E}_2/k_B\,T_H))\\
p=\text{Tr}[\vert e\rangle\langle e\vert\rho_{final}]=1/(1+exp(\mathcal{E}_1/k_B\,T_C))\,.
\end{eqnarray}
Therefore, the transition probability of the QOE can be written as
\begin{eqnarray}
\delta p=\frac{1}{(1+exp(\mathcal{E}_2/k_B\,T_H))}-\frac{1}{(1+exp(\mathcal{E}_1/k_B\,T_C))}\,.\label{dlpQOE}
\end{eqnarray}
Hence, for getting positive work, eq.(\ref{dlpQOE}) suggests that $\delta p>1$ which in turn leads to the condition
\begin{eqnarray}
T_H/\mathcal{E}_2>T_C/\mathcal{E}_1\label{Qcond}
\end{eqnarray}
which is much stronger than its classical analogue $T_H>T_C$.


\section{Results}\label{sec:Effects on transition probability and work output}
In this section we  analyse our findings for three different cases, namely, near boundary regime, intermediate boundary regime and far boundary regime of UQOE in the presence of a single reflecting boundary.

\subsection{Demarcation of regimes with respect to the parameters}

We first estimate the value of $z$ which depends on the acceleration of the qubit $(\alpha)$ and  $z_0$,  for different regimes. Studies in the context of trapped ultracold atoms \cite{Kozdon2018measuring} and superconducting circuits \cite{garcia2017entanglement} show that these quantum systems are effective to practically realize atom-field interactions due to accelerating qubits. In such systems taking the ultrafast variation of the qubit-field coupling, it has
been possible to achieve large acceleration up to $7\times 10^{17}m/s^2$ \cite{PhysRevB.92.064501}. In our analysis we choose the parameters mimicking the values for the above systems to a certain extent. Thus, during the first qubit-field interaction for a particular cycle, we take the value $\alpha=6\times 10^{17}m/s^2$, and note that in the intermediate boundary regime $\sin^2\left(\frac{\alpha\mathcal{T}}{2c}\right)\sim z^2$. Hence setting $\sin^2\left(\frac{\alpha\mathcal{T}}{2c}\right)=1= z^2$ fixes $\mathcal{T}=16\times10^{-9}s$ and $z_0=15\,cm$. Similarly, for the second qubit-field interaction in that same cycle we choose $\alpha=3\times 10^{17}m/s^2$. This fixes $\mathcal{T}=31\times10^{-9}s$ and $z_0=30\,cm$ in the intermediate boundary regime. With these choice of parameters, we now have three distinct regimes shown in Table \ref{table:parameter}.
\begin{table}[!htbp]
\begin{center}
\begin{tabular}{ |c|c|c|c|c|c| } 
 \hline
 Regime & $z$ & $z_{0H}$ (cm) & $z_{0C}$ (cm) \\ 
 \hline
 Near & less than 1 & less than 15 & less than 30\\ 
 \hline
 Intermediate & equal to 1 & equal to 15 &  equal to 30 \\ 
 \hline
 Far & greater than 1 & greater than 15 & greater than 30 \\
 \hline
\end{tabular}
\end{center}
\caption{Three different regimes in a particular thermodynamical cycle with $\alpha_{H}=6\times 10^{17}m/s^2$ and $\alpha_{C}=3\times 10^{17}m/s^2$.}
\label{table:parameter}
\end{table}\\

\subsection{Transition probability}
\begin{figure}[!htbp]
\begin{minipage}{.5\textwidth}
\centering
\includegraphics[width=0.8\linewidth]{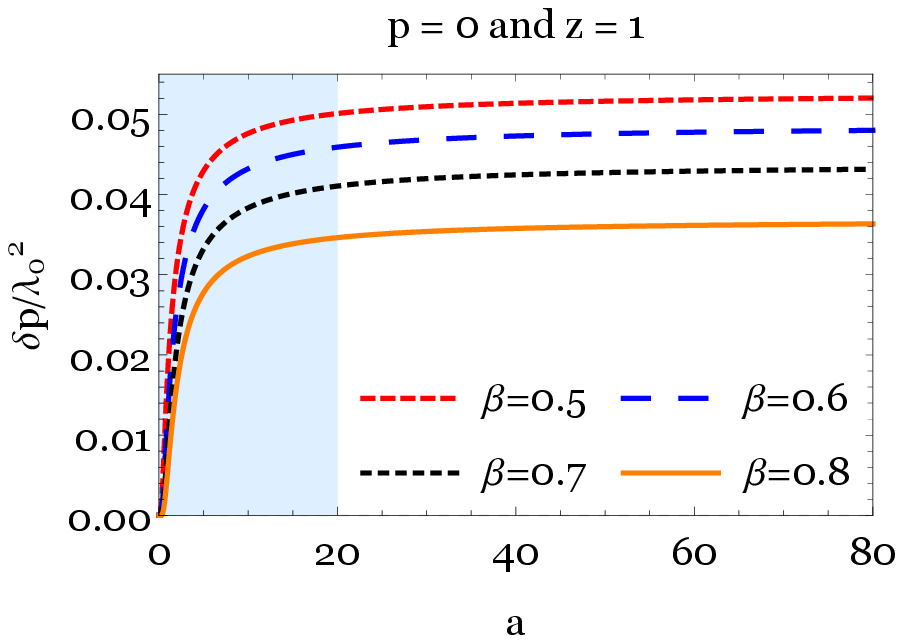}
\subcaption{Completely empty excited state.}\label{fig:sub-first1}
\end{minipage}
\begin{minipage}{.5\textwidth}
\centering
\includegraphics[width=0.8\linewidth]{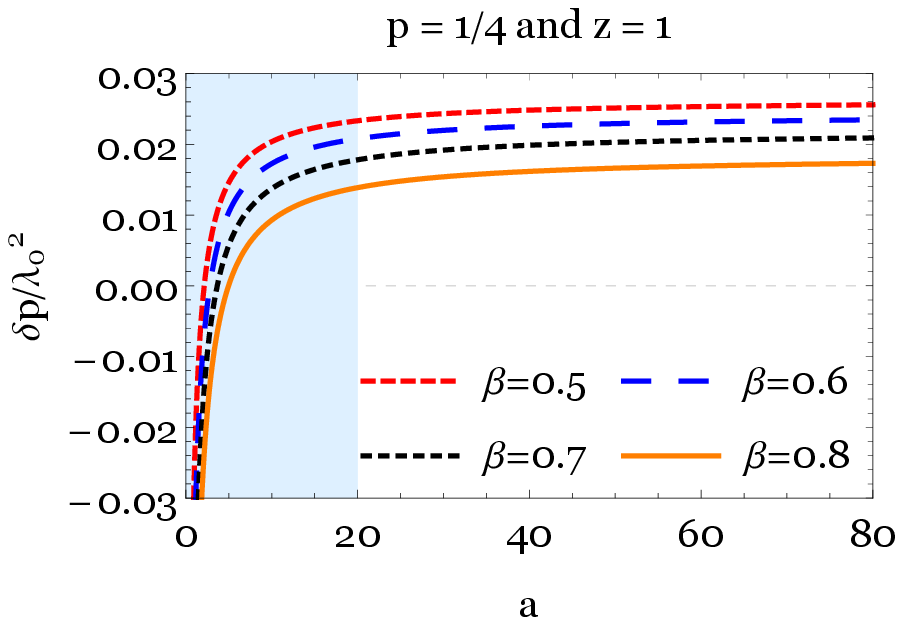}
\subcaption{Lowly populated excited state.}\label{fig:sub-second1}
\end{minipage}
\newline
\begin{minipage}{.5\textwidth}
\centering
\includegraphics[width=0.8\linewidth]{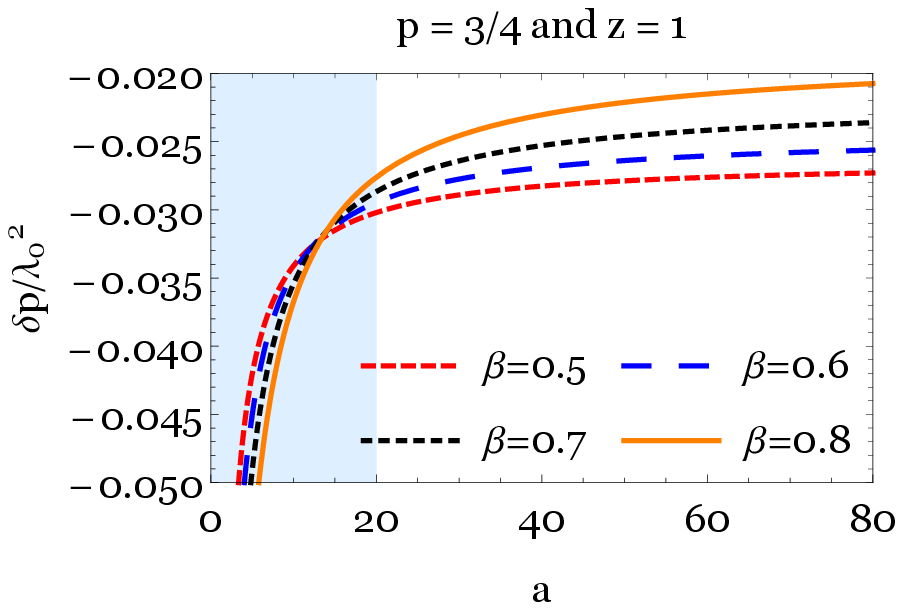}
\subcaption{Highly populated excited state.}
\label{fig:sub-third1}
\end{minipage}
\begin{minipage}{.5\textwidth}
\centering
\includegraphics[width=0.8\linewidth]{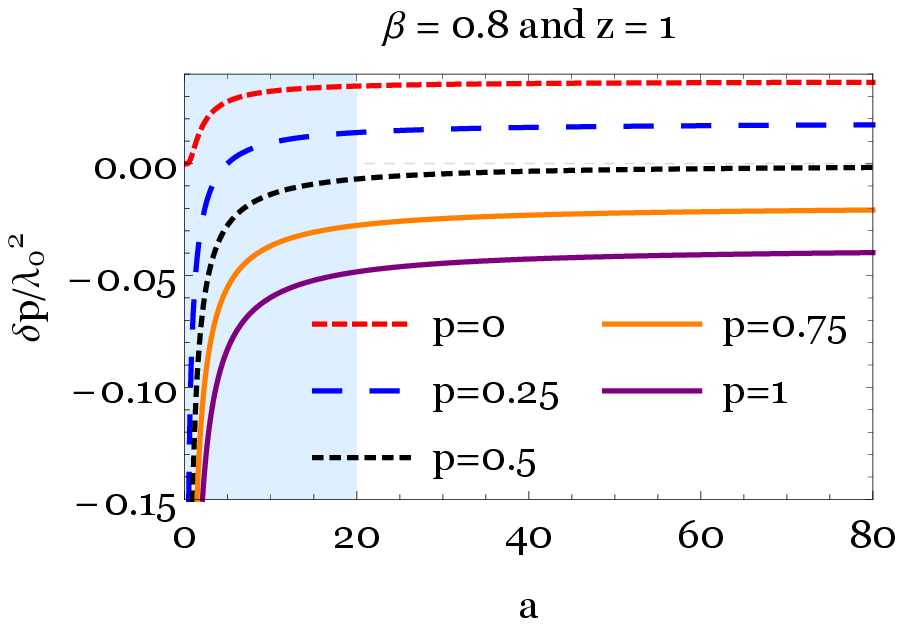}
\subcaption{Behaviour for $\beta=0.8$ and $z=1$.}
\label{fig:sub-fourth1}
\end{minipage}
\caption{Behaviour of the transition probability with respect to reduced acceleration $a$ in the intermediate boundary regime for various values of $p$ and $\beta$.}
\label{fig:fig1}
\end{figure}

\begin{figure}[!htbp]
\begin{minipage}{.5\textwidth}
\centering
\includegraphics[width=0.8\linewidth]{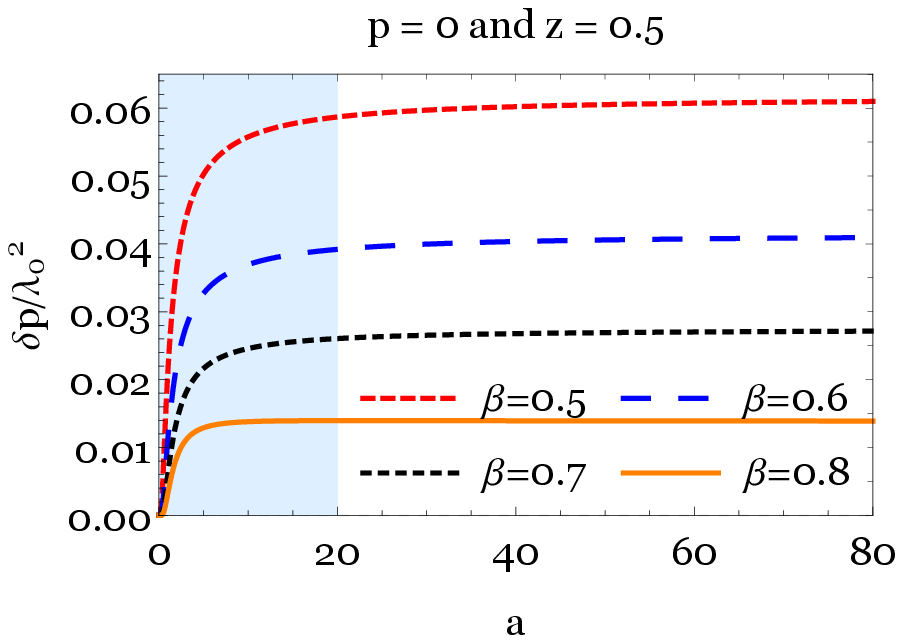}
\subcaption{Completely empty excited state.}\label{fig:sub-first2}
\end{minipage}
\begin{minipage}{.5\textwidth}
\centering
\includegraphics[width=0.8\linewidth]{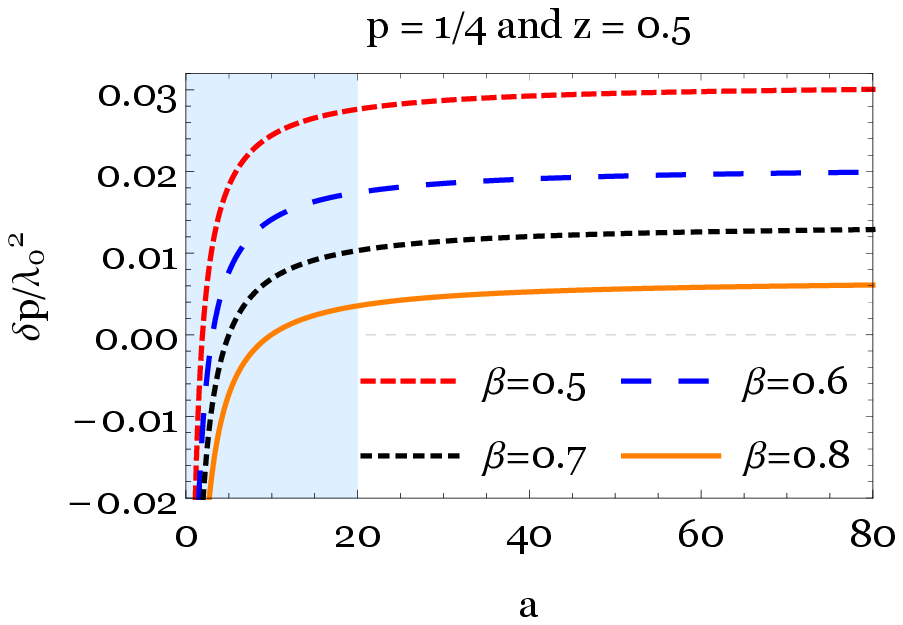}
\subcaption{Lowly populated excited state.}\label{fig:sub-second2}
\end{minipage}
\newline
\begin{minipage}{.5\textwidth}
\centering
\includegraphics[width=0.8\linewidth]{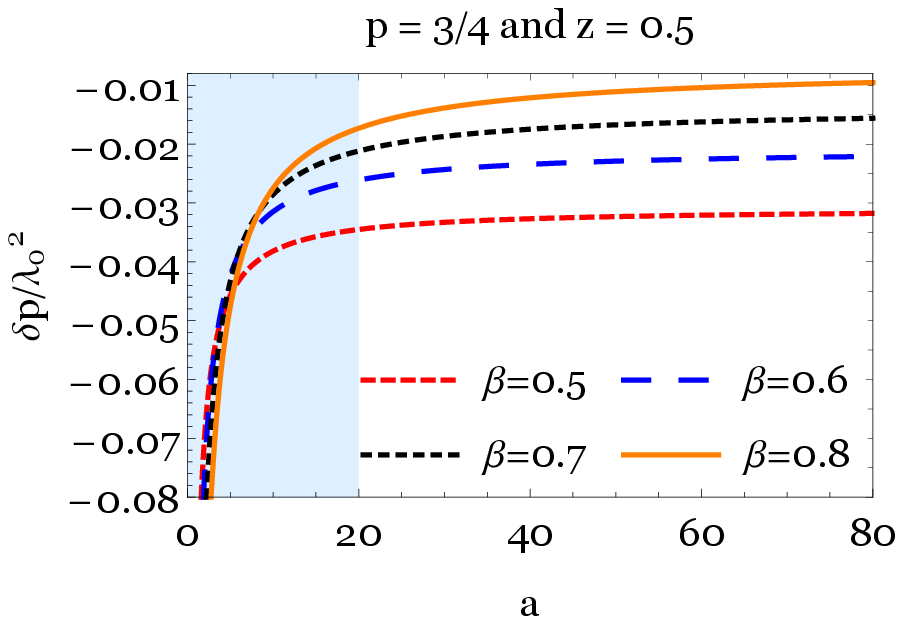}
\subcaption{Highly populated excited state.}
\label{fig:sub-third2}
\end{minipage}
\begin{minipage}{.5\textwidth}
\centering
\includegraphics[width=0.8\linewidth]{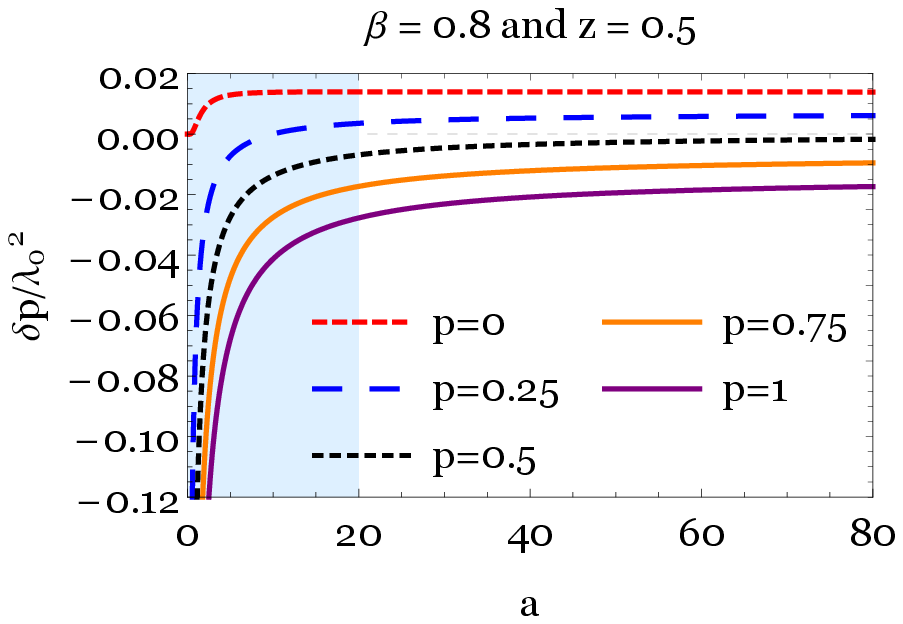}
\subcaption{Behaviour for $\beta=0.8$ and $z=0.5$.}
\label{fig:sub-fourth2}
\end{minipage}
\caption{Behaviour of the transition probability with respect to reduced acceleration $a$ in the near boundary regime for various values of $p$ and $\beta$.}
\label{fig:fig2}
\end{figure}

\begin{figure}[!htbp]
\begin{minipage}{.5\textwidth}
\centering
\includegraphics[width=0.8\linewidth]{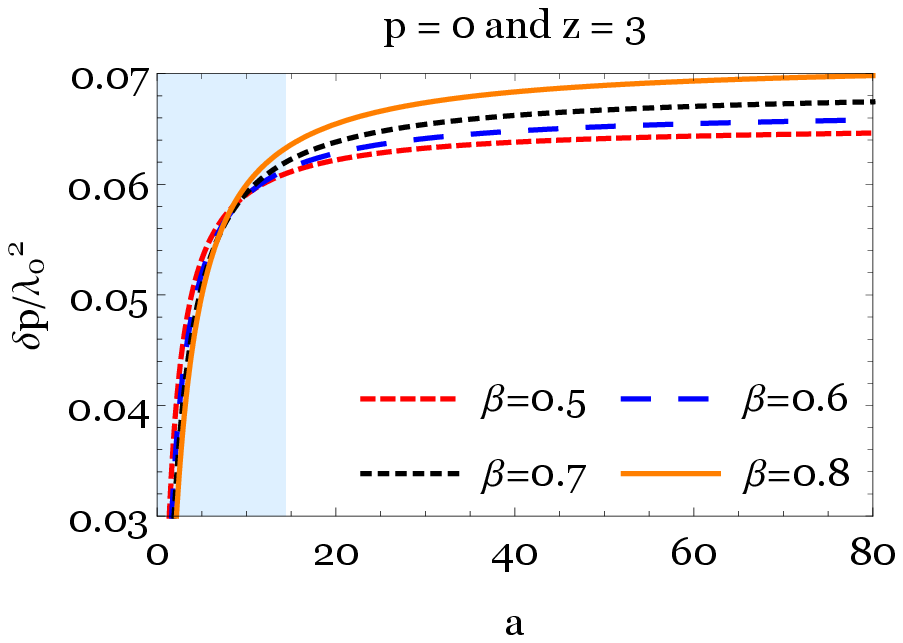}
\subcaption{Completely empty excited state.}\label{fig:sub-first3}
\end{minipage}
\begin{minipage}{.5\textwidth}
\centering
\includegraphics[width=0.8\linewidth]{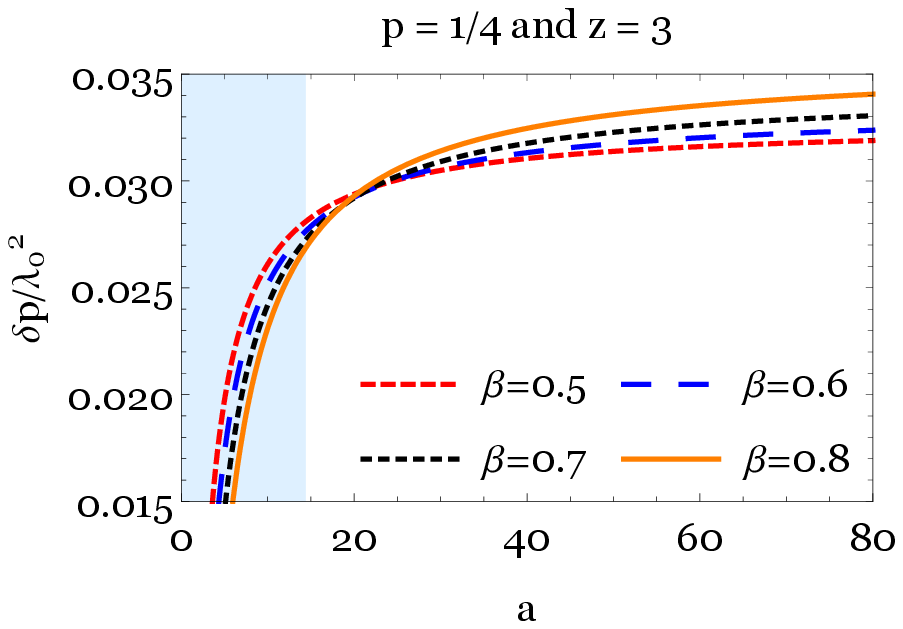}
\subcaption{Lowly populated excited state.}\label{fig:sub-second3}
\end{minipage}
\newline
\begin{minipage}{.5\textwidth}
\centering
\includegraphics[width=0.8\linewidth]{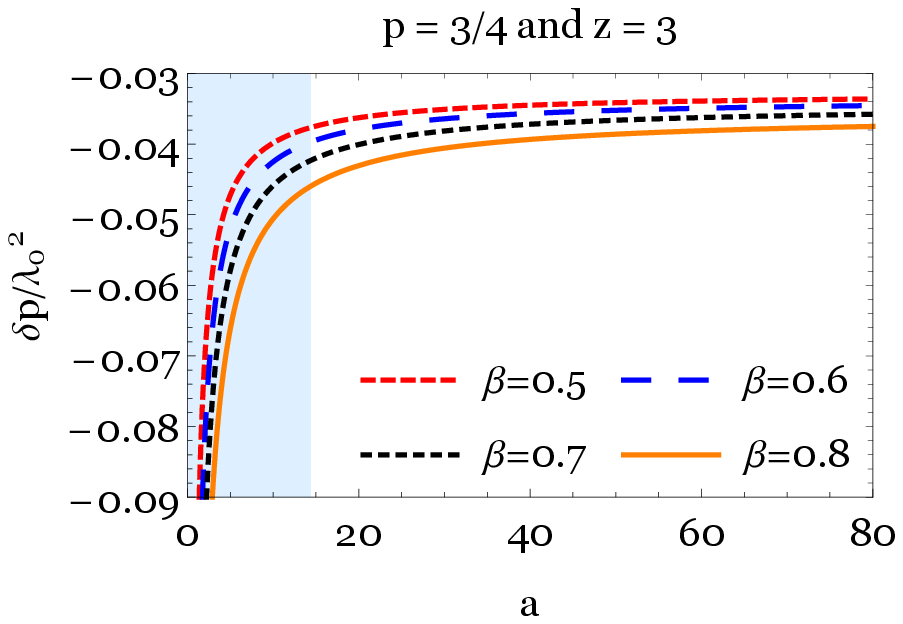}
\subcaption{Highly populated excited state.}
\label{fig:sub-third3}
\end{minipage}
\begin{minipage}{.5\textwidth}
\centering
\includegraphics[width=0.8\linewidth]{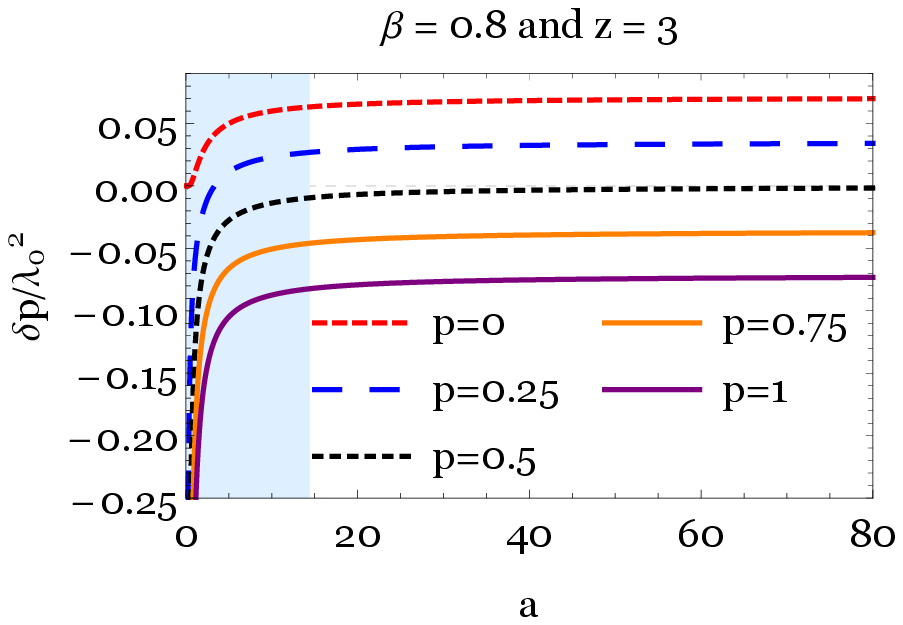}
\subcaption{Behaviour for $\beta=0.8$ and $z=3$.}
\label{fig:sub-fourth3}
\end{minipage}\caption{Behaviour of the transition probability with respect to reduced acceleration $a$ in the far boundary regime for various values of $p$ and $\beta$.}
\label{fig:fig3}
\end{figure}

We have already seen from our calculations, the transition probability $\delta p$ between two energy levels of the qubit depends on various parameters. Here we plot the behaviour of the transition probability between two energy levels of the qubit with respect to reduced acceleration $a$ in three different regimes, namely, intermediate, near and far boundary regime, respectively, for various values of $p,\,\beta$ and $z$ in Figures \ref{fig:fig1}, \ref{fig:fig2} and \ref{fig:fig3} respectively. Following the estimation given in Table \ref{table:parameter}, we consider $z=0.5$ for the near boundary regime, $z=1$ for intermediate boundary regime and $z=3$ for far boundary regime. For each regime we consider four qubit velocities $\beta=0.5, 0.6, 0.7,0.8$.

From Figure \ref{fig:fig1}, it is observed that in the intermediate boundary regime, the transition probability  corresponding to each qubit velocity $\beta$ ubiquitously increases with the increment of reduced acceleration, and from Unruh effect it is directly associated with the increment of temperature. The shaded blue region shows the region where the perturbative scheme breaks down for small $a$, and $\delta p$ diverges for $p\neq0$.

The transition probability eq.\eqref{deltap0}, for the case where the probability of a qubit being in the excited state is zero ($p=0$), is  plotted in Figure \ref{fig:sub-first1}. Here we observe that the transition probability $\delta p$ is always positive for all four chosen qubit velocities. Comparing with the UQOE without any reflecting boundary \cite{arias2018unruh, gray2018scalar}, we observe that for a particular value of reduced acceleration $a$, the transition probability decreases with the increase of parameter $\beta$. 

In Figure \ref{fig:sub-second1}, the transition probability for the case where the qubit has a lowly populated excited state ($p=1/4$) is plotted. As the excited state is lowly populated, the
value of $\delta p$ is lesser here. Divergence  occurs in the region where reduced acceleration is very low. In Figure \ref{fig:sub-third1}, an initially highly populated qubit state ($p=3/4$) is plotted. In this case $\delta p$ is negative, which indicates that due to  a less populated ground state after the interaction of the quantum field, a de-excitation process occurs. In Figure \ref{fig:sub-fourth1}, different initial excitation probabilities for a fixed value of $\beta$ and $z$ are plotted. Here we get positive transition probability $\delta p$ for the value $p=0$ for all values of $a$. For $0<p<1/2$, $\delta p$ is initially negative for lower values of $a$ and becomes positive for higher values of $a$. However we get negative $\delta p$ for the value $1>p>1/2$ for all values of $a$. For the critical value $p=1/2$, $\delta p$ is initially negative and approaches to zero as the value of $a$ increases.

We now display our results for the near boundary regime.  From the Figures, it is also observed that for a particular value of $\beta$, $\delta p$  continuously increases with the increment of reduced acceleration. For small values of reduced acceleration $a$,  the perturbative scheme breaks down and $\delta p$ diverges for all $p\neq0$ cases. From Figure \ref{fig:sub-first2}, it is seen that for the case where the probability of a qubit being in the excited state is zero ($p=0$), $\delta p$ remains positive but due to the effect of the reflecting boundary, it is observed that as the distance between the reflecting boundary and the qubit decreases compared to the intermediate boundary regime for a particular reduced acceleration, the transition probability for higher qubit velocity decreases rapidly compared to the intermediate boundary regime. From Figure \ref{fig:sub-second2}, we again observe similar behaviour (as the intermediate boundary regime) of the transition probability for the lowly populated excited state of the qubit ($p=1/4$).
In Figure \ref{fig:sub-third2}, transition probability corresponding to an initially highly populated qubit state, $p=3/4$ is plotted. In this case $\delta p$ is also negative in nature.
In Figure \ref{fig:sub-fourth2}, different initial excitation probabilities for a fixed value of $\beta$ are plotted. Here  we gain observe similar behavoiur with Figure \ref{fig:sub-fourth1}.

Next, considering the far boundary regime, from Figure \ref{fig:fig3}, it is observed that the nature of the Figure \ref{fig:sub-first3}, \ref{fig:sub-second3} and \ref{fig:sub-third3} are 
qualitatively similar to the Figure \ref{fig:sub-first1}, \ref{fig:sub-second1} and \ref{fig:sub-third1}, except that for a particular value of reduced acceleration, the transition probability increases with increase in the qubit velocity, which happens in the case of UQOE without any reflecting boundary \cite{arias2018unruh, gray2018scalar}. Here we once again observe that in the far boundary limit $\delta p$ remains always positive for the case where the probability of a qubit being in the excited state is zero ($p=0$). Here too,  Figure \ref{fig:sub-fourth3} shows the variation of $\delta p$ with different initial excitation probabilities for a fixed value of $\beta$. As expected,
the far boundary regime yields results that approach the case of UQOE without any reflecting boundary \cite{arias2018unruh, gray2018scalar}.

The impact of the reflecting boundary is further clearly revealed through a comparison between the behaviour of the transition probability with respect to reduced acceleration $a$ in the intermediate, near and far boundary regimes, as plotted in Figures (\ref{fig:sub-first1}, \ref{fig:sub-first2}, \ref{fig:sub-first3}), respectively. It can be seen that when the reflecting boundary is close to the qubit, for a fixed value of reduced acceleration the transition probability decreases with the increase of parameter $\beta$.  In the intermediate regime, it is observed that for a fixed value of reduced acceleration, the transition probability still decreases with the increase of parameter $\beta$, but the transition probability corresponding to the higher qubit velocity is greater than the corresponding value in the near boundary limit. Interestingly, shifting the reflecting boundary further, we find that in the far boundary limit, for a fixed value of reduced acceleration, the transition probability increases with the increase of parameter $\beta$. However, for  fixed values of $p$, $\beta$ and $a$, the transition probability increases when the distance between the reflecting boundary and the qubit is increased. This occurs because, as the distance between the boundary and the qubit is increased, more number of field modes take part in the interaction between the scalar field and the qubit, which in turn increases the transition probability. Hence, the behaviour of transition probability shows the effect of the reflecting boundary clearly.

\subsection{Work output in the presence of a reflecting boundary}
In this subsection we  calculate the work output of the UQOE in the presence of a reflecting boundary. We have seen earlier that presence of the reflecting boundary does not affect qubit's thermodynamical steps. To calculate the work output of the UQOE, at first we have to ensure that the thermodynamical cycle is closed. In order  to achieve the cyclicity of the UQOE in presence of the boundary, we employ the constraint condition,
\begin{eqnarray}
\delta p_H(a_H, p, \beta)+\delta p_C(a_C, p, \beta)=0 \,.\label{const}
\end{eqnarray}
Substituting eq.\eqref{deltap0} in the above constraint condition eq.\eqref{const}, we get
\begin{eqnarray}
(1-2p)\,\bigg[ \mathcal{F}\left(\frac{1}{a_H},\, 2\tanh^{-1}(\beta)\right)+\mathcal{F}\left(\frac{1}{a_C},\, 2\tanh^{-1}(\beta)\right)\bigg] &=& p\,\bigg[\Delta\mathcal{F}\left(\frac{1}{a_H},\, 2\tanh^{-1}(\beta)\right)\bigg.\nonumber\\
&+&\bigg.\Delta\mathcal{F}\left(\frac{1}{a_C},\, 2\tanh^{-1}(\beta)\right)\bigg]\,.\nonumber\\
&\,&
\end{eqnarray}
Therefore
\begin{eqnarray}
\frac{p}{1-2p}=\frac{\mathcal{F}\left(\frac{1}{a_H},\, 2\tanh^{-1}(\beta)\right)+\mathcal{F}\left(\frac{1}{a_C},\, 2\tanh^{-1}(\beta)\right)}{\Delta\mathcal{F}\left(\frac{1}{a_H},\, 2\tanh^{-1}(\beta)\right)+\Delta\mathcal{F}\left(\frac{1}{a_C},\, 2\tanh^{-1}(\beta)\right)}\,.
\end{eqnarray}
\begin{figure}[!htbp]
\centering
\includegraphics[width=0.8\linewidth]{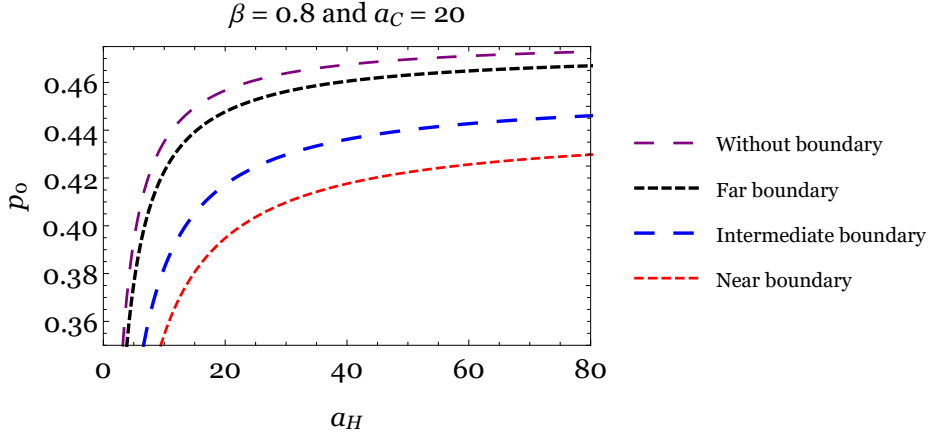}
\caption{Behaviour of the critical probability $p_0$ with respect to $a_H$ for fixed values of $\beta$ and $a_C$.}
\label{fig:p0}
\end{figure}\\
Defining $\mathcal{P}$ as
\begin{eqnarray}
\mathcal{P}\equiv\frac{\mathcal{F}\left(\frac{1}{a_H},\, 2\tanh^{-1}(\beta)\right)+\mathcal{F}\left(\frac{1}{a_C},\, 2\tanh^{-1}(\beta)\right)}{\Delta\mathcal{F}\left(\frac{1}{a_H},\, 2\tanh^{-1}(\beta)\right)+\Delta\mathcal{F}\left(\frac{1}{a_C},\, 2\tanh^{-1}(\beta)\right)}.
\end{eqnarray}
the initial population of the excited state of the qubit turns out to be
\begin{eqnarray}
p=\frac{\mathcal{P}}{1+2\mathcal{P}}\,.
\end{eqnarray}
We define this initial excited state probability $p$ as the critical probability $p_0$, to ensure that the vacuum acts as a hot and cold reservoir,  at the same value of $p_0$ for the qubit with reduced acceleration $a_H$ and $a_C$, respectively. The closure of the cycle is ensured and the qubit returns to its initial state  for this particular value. From the thermodynamical analysis, we have already seen that during the first interaction with the vacuum, the qubit absorbs heat from the vacuum fluctuation and $\delta p(a_H, p, \beta)>0$. From figure(s)(\ref{fig:fig1}, \ref{fig:fig2}, \ref{fig:fig3}), it is seen that $\delta p$ is positive only when initial excited state probability $p=0$.

\noindent The behaviour of the critical excitation probability $p_0$ with respect to $a_H$ is displayed in Figure \ref{fig:p0}  for fixed values of $\beta=0.8$ and $a_C=20$. From the figure it is observed that the reflecting boundary reduces the critical probability of the qubit $p_0$. For a fixed value of $a_H$, we find that when the reflecting boundary is at a near distance, the critical probability of the excited state of the qubit having velocity $\beta=0.8$ is minimum and it gradually increases with increase in the boundary distance. In the presence of the boundary, the critical probability $p_0$ is also found to be bounded from both ends, i.e., $0\leq p_0<1/2$, which is consistent with the Figure(s)(\ref{fig:fig1}, \ref{fig:fig2}, \ref{fig:fig3}). Now, using the value of $p_0$, we can recast the transition probability of the qubit $\delta \bar{p}_H$ as
\begin{eqnarray}
\delta \bar{p}_H=\delta p(a_H, p_0, \beta),
\end{eqnarray}
which automatically ensures that the cyclic condition is satisfied. 

\noindent Next, considering eq.(\ref{netw}) we can write down the amount of total work done by the UQOE in presence of the reflecting boundary as
\begin{eqnarray}
W\equiv\langle W_{ext}\rangle=(\mathcal{E}_2-\mathcal{E}_1)\delta \bar{p}_H.
\end{eqnarray}
\begin{figure}[!htbp]
\centering
\includegraphics[width=0.8\linewidth]{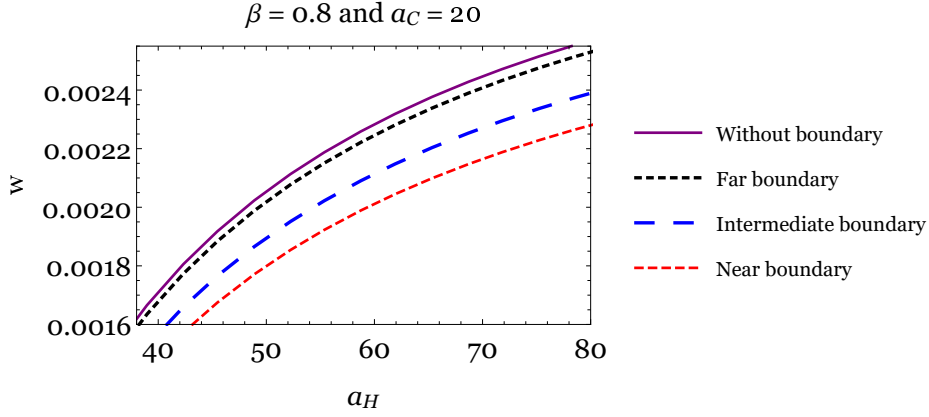}
\caption{Behaviour of the work output with respect to $a_H$ for fixed values of $\beta$ and $a_C$.}
\label{fig:work}
\end{figure}

\noindent In Figure \ref{fig:work}, we plot the amount of output work as a function of $a_H$ for  fixed values of $\beta=0.8$ and $a_C=20$ by taking $\mathcal{E}_2-\mathcal{E}_1=1$. From the figure it is observed that in our model of relativistic quantum thermal heat engine, the work output depends on the position of the reflecting boundary. Comparing with the work output of the UQOE \cite{arias2018unruh}, we observe that in the near mirror limit the work output is minimum, and it gradually increases with increase in the distance between the boundary and the qubit. This is consistent with the fact that the critical probability $p_0$ is a monotonic function with respect to the location of the boundary (see Figure \ref{fig:p0}). We also observe that when $a_H>a_C$, the cycle behaves as a thermal machine and positive work is done. In order to act as a thermal mechine, the system absorbs heat from the hot quantum vacuum $\delta \bar{p}_H>0$, which in turn implies the condition,
\begin{eqnarray}
\alpha_H/\mathcal{E}_2>\alpha_C/\mathcal{E}_1.
\end{eqnarray}
Therefore, we also find a similar condition for UQOE in the presence of a reflecting boundary, which is the same as eq.(\ref{Qcond}), and  is stronger compared to its classical counterpart.

\noindent The net work done by the engine and total amount of heat transfer between the qubit and the quantum vacuum in the presence of the reflecting boundary is
given by eq(s).(\ref{netw}, \ref{neth}). We have found that the values of
these quantities satisfy the energy conservation principle. From the
expressions  eq(s).(\ref{netw}, \ref{neth}), we can calculate the efficiency of the heat engine. Since we have employed external stimulation to change the energy gap of the qubit levels during the adiabatic expansion and contraction discussed in section \ref{sec:Th-Analysis}, therefore, the amount of work done by the qubit is  $\langle W_{\text{ext}}\rangle=-\langle W\rangle  $. Hence, the efficiency is gievn by
\begin{eqnarray}
\eta=\frac{\langle W_{\text{ext}}\rangle }{Q_2}=1-\frac{\mathcal{E}_1}{\mathcal{E}_2}\label{eff}
\end{eqnarray}
It may be noted that as $Q_2$ is also proportional to  $\delta \bar{p}_H$, therefore, the amount of absorbed heat is also reduced in the presence of a reflecting boundary. Hence, from eq.(\ref{eff}) it is observed that the efficiency in our model in independent of any boundary effect. It only depends on the energy gap ratio between the two levels of the qubit and takes a form identical to the UQOE without any boundary \cite{arias2018unruh, gray2018scalar, xu2020unruh}.


\section{Conclusions}\label{sec:Con}

In this paper, we have proposed a  model for the relativistic quantum analogue of the classical Otto heat engine.  In our model, a uniformly accelerated qubit (Unruh-DeWitt detector) acts as the working substance, and is coupled to a massless quantum scalar field in the presence of a perfectly reflecting boundary which obeys the Dirichlet boundary condition. The reflecting boundary is the new ingredient that we introduce in our work, which and has important physical consequences. Using the notion of the Unruh effect, the quantum vacuum behaves as a thermal bath, and we have uncovered certain interesting features associated
with the process of work extraction from the quantum vacuum fluctuations of a  quantum scalar field in presebce of a reflecting boundary. 

It is has been observed earlier that the correlation function between  scalar fields, commonly known as the Wightman function, gets significantly modified by the presence of a reflecting boundary \cite{rizzuto2007casimir}. Since the response function of the qubit depends on this correlation function  \cite{sachs2017entanglement, louko2016unruh, takagi1986vacuum, PhysRevD.93.024019}, an extra contribution due to the presence of the reflecting boundary appears  \cite{arias2018unruh, gray2018scalar}. From the structure of this correlation function, we find that three different cases emerge, i.e., the near boundary regime, the intermediate boundary regime, and the far boundary regime. 
We show that the near boundary case is the one where the role of the reflecting boundary is 
most prominently felt,  and the far boundary limit corresponds to the case in which one can smoothly go to the case where the boundary is absent. Choosing 
experimentally realizable values of the qubit acceleration and the distance between the qubit and the reflecting boundary, we estimate a parameter which determines the applicability of each approximation limit.

Our analysis leads to several interesting results.
We find that when the reflecting boundary is close to the qubit, for a 
fixed value of qubit acceleration the transition probability decreases with the increase of the qubit velocity. In the intermediate limit, it is observed that  the transition probability corresponding to a higher qubit velocity is greater than the corresponding value in the near boundary limit. Shifting the  boundary further, we find that in the far boundary limit the transition probability starts increasing with increase in the qubit velocity for a particular value of qubit acceleration.  The  effect of the reflecting boundary is clearly 
manifested through the behaviour of the atomic transition probability. 

We further observe that the reflecting boundary reduces the critical probability of the qubit compared to the Unruh quantum Otto engines (UQOE) \cite{arias2018unruh} without a boundary. For a fixed value of $a_H$, we find that when the reflecting boundary is at a near distance, the critical probability of the excited state of the qubit is minimum and it gradually increases with increase in the distance between the reflecting boundary and the qubit. This reveals that the critical probability is a monotonically increasing function of the distance between the reflecting boundary and the qubit. Next,
comparing the work output of this new model with the usual UQOE \cite{arias2018unruh}, we find that the work extraction gets inhibited due to the presence of the reflecting boundary. It is observed that the output work of this quantum relativistic heat engine depends on the position of the reflecting boundary, and maximum work output is obtained when the distance between the qubit and the reflecting boundary is maximum. Furthermore, we also observe that the work output is also a monotonically increasing function of the distance between the reflecting boundary and the qubit. This is compatible with the result obtained for the critical excitation probability. However, for the entire cycle, it is observed that the efficiency of our engine is identical to that of the usual UQOE. 

Our approach opens up several new directions for further studies. First, 
 our  model of relativistic quantum thermal machine can be probed further without the linearized approximation by considering higher order of interactions. Secondly, it would be interesting to explore whether the boundary effects on relativistic heat engines may be reversed using fermionic quantum fields. Finally, this work can also be extended to 
 the domain of an experimental   superconducting cavity setup \cite{del2022quantum} in order to verify the boundary effects on the efficiency and work output of the model proposed here.


\section*{Acknowledgement}
AM and ASM acknowledges support from project no. DST/ICPS/QuEST/2019/Q79 of the Department of Science and Technology (DST), Government of India. The authors would also like to thank the referees for very useful comments and suggestions.


\bibliographystyle{JHEP.bst}
\bibliography{Reference.bib}
\end{document}